\documentclass[fleqn,10pt,iop]{emulateapj}

\usepackage{url}
\usepackage{multirow}
\usepackage{amsmath}
\usepackage{xcolor}
\usepackage{nccmath}

\newcommand{\logL}{\log\mathcal{L}}
\newcommand{\unit}[1]{\footnotesize #1}
\newcommand{\PAPER}{\mathrm{PAPER}}
\newcommand{\Nconf}{31}
\newcommand{\Nsrc}{32}
\definecolor{orange}{RGB}{255,127,0}

\slugcomment{Accepted to ApJ}

\shorttitle{Southern Hemisphere Flux Scale}
\shortauthors{Jacobs et al.}

\begin{document}

\title{A Flux Scale for Southern Hemisphere 21cm EoR Experiments}
\author{
Daniel C. Jacobs\altaffilmark{1},
Aaron R. Parsons\altaffilmark{2,8},
James E. Aguirre\altaffilmark{3},
Zaki Ali\altaffilmark{2},
Judd Bowman\altaffilmark{1},
Richard F. Bradley\altaffilmark{4,5,6},
Chris L.  Carilli\altaffilmark{7},
David R. DeBoer\altaffilmark{8},
Matthew R. Dexter\altaffilmark{8},
Nicole E. Gugliucci\altaffilmark{5},
Pat Klima\altaffilmark{5},
Dave H. E. MacMahon\altaffilmark{8}
Jason R. Manley\altaffilmark{9},
David F. Moore\altaffilmark{3},
Jonathan C. Pober\altaffilmark{2},
Irina I. Stefan\altaffilmark{10},
William P. Walbrugh\altaffilmark{9}}

\altaffiltext{1}{School of Earth and Space Exploration, Arizona State U., Tempe, AZ}
\altaffiltext{2}{Astronomy Dept., U. California, Berkeley, CA}
\altaffiltext{3}{Dept. of Physics and Astronomy, U. Pennsylvania, Philadelphia, PA}
\altaffiltext{4}{Dept. of Electrical and Computer Engineering, U. Virginia, Charlottesville, VA}
\altaffiltext{5}{National Radio Astronomy Obs., Charlottesville, VA}
\altaffiltext{6}{Dept. of Astronomy, U. Virginia, Charlottesville, VA}
\altaffiltext{7}{National Radio Astronomy Obs., Socorro, NM}
\altaffiltext{8}{Radio Astronomy Lab., U. California, Berkeley, CA}
\altaffiltext{9}{Square Kilometer Array, South Africa Project, Cape Town, South Africa}
\altaffiltext{10}{Cavendish Lab., Cambridge, UK}

\begin{abstract}
We present a catalog of spectral measurements covering a 100--200-MHz band for 32 sources, derived
from observations with 
a 64-antenna deployment of the 
Donald C. Backer Precision Array for Probing
the Epoch of Reionization (PAPER) in South Africa.
For transit telescopes such as PAPER, calibration of the primary beam is
a difficult endeavor, and errors in this calibration are a major source of error in
the determination of source spectra.
In order to decrease reliance on accurate beam calibration,
we focus on calibrating sources in a narrow declination range
from -46\arcdeg{} to -40\arcdeg{}.
Since sources at similar declinations follow nearly identical paths through the primary beam, this
restriction greatly reduces errors associated with beam calibration,
yielding a dramatic improvement in the accuracy of derived source spectra.
Extrapolating from higher frequency
catalogs, we derive the flux scale using a Monte-Carlo fit across multiple sources that 
includes uncertainty from both catalog and measurement errors. Fitting spectral 
models to catalog data and these new PAPER measurements, we derive new flux models for
Pictor A and \Nconf{} other sources at nearby declinations, 
90\% are found to confirm and refine a power-law model for flux density.
Of particular importance is the new Pictor A flux model, which is accurate to
1.4\% and shows, in contrast to previous models, that between 100 MHz and 2 GHz, the spectrum of Pictor A 
is consistent with a single power law given by a flux at 150 MHz of
382$\pm$5.4 Jy, and a spectral index of -0.76$\pm$0.01.  
This accuracy represents 
an order of magnitude improvement over previous measurements in this band, and
is limited by the uncertainty in the catalog measurements used
to estimate the absolute flux scale.  The simplicity and improved accuracy of Pictor A's spectrum
make it an excellent calibrator in a band of
importance to experiments seeking to
measure 21cm emission from the Epoch of Reionization. 
\end{abstract}


\section{Introduction}

Numerous radio telescopes are now exploring the prospects for using
measurements of highly redshifted 21cm emission to inform our understanding of
cosmic reionization in the redshift range $z>6$, corresponding to radio
frequencies below 200 MHz (see reviews in
\citealt{Furlanetto:2006p2267,Morales:2010p8093,Pritchard:2012p9555}).  These
include telescopes aiming to measure the global temperature change of 21cm
emission during the Epoch of Reionization (EoR), such as the Compact
Reionization Experiment (CoRE), the Zero-spacing Interferometer \citep{Raghunathan:2011}  and the Experiment to Detect the Global EoR
Signature (EDGES; \citealt{Bowman:2010p8546}), and interferometers aiming to
measure the power spectrum of 21cm EoR emission, such as the Giant Metre-wave
Radio Telescope (GMRT;
\citealt{Paciga:2011p9470,Paciga:2013p9943})\footnote{\url{http://gmrt.ncra.tifr.res.in/}},
the LOw Frequency ARray (LOFAR;
\citealt{Yatawatta:2013p9699,Haarlem:2013p10011})\footnote{\url{http://www.lofar.org/}}, the
Murchison Widefield Array (MWA;  \citealt{Bowman:2013p9950};
\citealt{Tingay:2013p9022}))\footnote{\url{http://www.mwatelescope.org/}}, and
the Donald C. Backer Precision Array for Probing the Epoch of Reionization
(PAPER; \citealt{Parsons:2010p6757,Pober:2013p9942})\footnote{\url{http://eor.berkeley.edu/}}.

Given the immense science potential in detecting 21cm emission from the EoR,
a great deal of research has focused on measuring the spectral and spatial variation of foreground emission
in the 100-200~MHz band ($z=6-13$ in the 21cm line), which dwarfs the 21cm signal by orders of magnitude
\citep{Furlanetto:2006p2267}. 
In particular, the spectral properties of extra-galactic point-sources are important
both because they are valuable calibration references
and because they are strong foreground emitters that must be removed from 21cm EoR measurements.
With the sparse availability of measured foreground properties in the
100--200-MHz frequency band over large areas of the sky \citep{deOliveiraCosta:2008p2242}, continued
foreground characterization is a
vital step en route to any 21cm EoR detection.
At these low
frequencies, the Southern sky is much less well known than the North;
catalog source fluxes at 150MHz are inaccurate at the 20\% level
for $\delta<-20$\arcdeg{} \citep{Slee:1995p7541,Vollmer:2005p6425}.
Both PAPER and the MWA are located in the southern hemisphere at radio-quiet
reserves being prepared for the upcoming Square Kilometre Array, and hence,
the
most extensive surveying work is now being conducted by the EoR experiments
themselves \citep{Jacobs:2011p8438,Williams:2012p8768,Bernardi:2013p9859}.

One significant complication to improving the state of affairs in foreground characterization
is that many 21cm EoR experiments, including LOFAR in the northern hemisphere, and
PAPER and MWA in the southern hemisphere, are designed for drift-scan observations or steered via phased array.
This design decision has largely been driven by the simplicity and cost-effectiveness of
phased and/or correlated dipoles to achieve the aggressive sensitivity requirements
for measuring the 21cm power spectrum of reionization \citep{Parsons:2012p9028,Beardsley:2013p9952,Jelic:2008p2130}.
Adding to the challenge, these telescopes 
cover much wider fields of view ($>10$\arcdeg) and bandwidths
($\sim 100$\% fractional) than traditional dish telescopes.
Because they do not physically point, flux calibration for such arrays relies heavily
on an accurate model of the primary beam response to correct for an
the apparent flux scale that varies across the sky.
This direction dependent gain is currently uncertain to 10\% or higher
\citep{Pober:2012p8800}, and comprises a large fraction of the 20\% flux uncertainty between current telescopes \citep{Jacobs:2013p9908}.

In this paper, we set out to significantly improve the accuracy of spectral measurements between 
100--200 MHz for a set of bright sources in the declination range -46\arcdeg{} to -40\arcdeg{} that
are of particular value for southern-hemisphere 21cm EoR experiments such as PAPER and the MWA.
Using the fact that, for this restricted declination range,
sources transit through a nearly identical primary beam response pattern, we are able to avoid one of the
most debilitating source of error in these measurements: the primary beam.

In \S\ref{sec:background} we provide some background on uncertainty in early EoR-band catalogs,
and explain our choice of calibrators.
In \S\ref{sec:approach} we describe our approach for measuring source spectra with
drift-scan observations and deriving an absolute flux scale from catalog data.  
In \S\ref{sec:Observations} we detail the instrumental 
setup, observations and analysis method followed. Sections \ref{sec:mcmc} through
\ref{fig:gain} detail our approach for fitting a global flux scale and spectral models for each source.
We use these fits in \S\ref{sec:results} to understand how well PAPER data agree
with previous measurements and conclude in \S\ref{sec:Conclusion}.

\section{Background}
\label{sec:background}

Historically, the best Southern Hemisphere 
EoR band data were by \citet{Slee:1995p7541} with Culgoora Circular
Array\footnote{Known during daylight hours as Culgoora Radio Heliograph} and
various higher frequency measurements with Parkes.  These data are typically
uncertain to 20\% or higher and provide little coverage of the EoR band beyond
a single narrow-band data point. 
More recent surveys include narrow-band surveys by the GMRT \footnote{\url{http://tgss.ncra.tifr.res.in/}} and Mauritius \citep{Pandey:2005p8687}, a
deep survey of the region near Hydra A by the 32 antenna MWA prototype
\citep{Williams:2012p8768} and a wide field survey by PAPER, also with 32
elements \citep{Jacobs:2011p8438}. Several sub-channels were provided in the
Williams catalog, though with 60-80\% error bars ---large compared to the 30\%
uncertainty on their wide band measurements.  The latter cover the band and
spatial scales relevant to EoR measurements but are limited by the accuracy of
the primary beam \citep{Jacobs:2013p9908} as well as the lack of precise in-band flux calibrators.

The response of the primary beam is of critical importance to EoR measurements.
Differences between the polarization responses cause leakage of polarized
signals into the total intensity measurement possibly corrupting the 
EoR power spectrum \citep{Moore:2013p9941}.  The primary beam 
shape is also critical to measuring and subtracting foregrounds 
\citep{Bernardi:2013p9859,Sullivan:2012p9457,Morales:2012p8790} a process which, to be effective, must be done to better than 1\%
precision for the brightest sources \citet{Liu:2009p4762,Bowman:2009p7816} .
A method for decoupling uncertain fluxes from the uncertain beam has been
described by \citet{Pober:2012p8800}. In simulation the method was able to
achieve 3 to 10\% accuracy in measuring the primary beam, depending on the
number of antennae and other variables. In particular, it emphasized the need
for many repeated measurements of each alt-az pointing, which in that
case were found by assuming 180\arcdeg{} symmetry. Further investigation
is under way to improve and implement this method and would be greatly aided by
the availability of precise flux measurements unaffected by primary beam
uncertainty. 

EoR measurements by PAPER and the MWA in the southern hemisphere have focused on the coldest regions where
galactic foregrounds are minimal, with the majority of possible observing time falling around
RA=4h,Dec=-30. The brightest and least-resolved calibrator in this region is Pictor A 
(5h19m49.1,-45d46m45.0). Pictor A is a nearby FR-II type radio galaxy 
 similar to Cygnus A.  At $\sim$400Jy, Pictor is bright and sufficiently distant from other 
bright sources to make it eminently suitable as both a phase and flux calibrator. Its apparent
size of $\sim$8\arcmin{} is smaller than the scales being probed by current EoR instruments, 
making it suitable for precision calibration, with only a modest
level of resolution effects.
However, 
like most other sources, precise flux measurements in the EoR band are not available.
The previous best EoR band measurement
is uncertain to 12\% and appears to imply spectral flattening in the EoR band
\citep{Perley:1997p9312}. 
 

Establishing an accurate spectrum for Pictor A is of particular importance for
PAPER --- a dedicated EoR experiment 
that employs drift-scanning, dual-polarization dipole antennas 
tuned for efficient operation over a 120--170-MHz band.  PAPER is located in the South African Karoo desert
on the Square Kilometer Array
South Africa (SKA-SA) reserve, 100km north of the small town of Carnarvon.
The PAPER array has grown from 16 elements deployed in early 2009 to a
64-element imaging array in 2011 (see Figure \ref{fig:antpos}). 
Since November 2011 it has been arranged in a maximally redundant grid 
configuration to make deep power spectral integrations \citep{Parsons:2012p9028}.
Though highly sensitive as a power spectrum instrument, the maximally redundant array 
has a broad point spread function in the image domain. This severely limits the number 
of sources which may be used for flux calibration. Drift scanning across the sky with a 45\arcdeg{} FWHM
primary beam, there are very few unresolved, bright sources which are far from the galactic plane. 
Pictor A is bright and well enough separated from other emission to dominate the visibilities for a good fraction of
the EoR observing season, making it a desirable source to use for flux calibration.

\section{Approach}
\label{sec:approach}
In this section, we describe our general approach to controlling the impact of beam model errors on measured
spectra derived from 
drift-scan observations 
of Pictor A and a selection of known, bright sources.  Our approach uses a set of ``source tracks'' as
a function of frequency and time.  Each source track is a beam, formed by phasing measured visibilities toward
known source locations as they drift through the primary beam, and summing over antenna pairs, as
described in \S\ref{sec:beamforming}.  Since sources fall
at different positions
within the primary beam, it is generally not possible to relate the fluxes of sources at different positions to
one another without an accurate beam model.  To date, the accuracy of source flux measurements in the
southern hemisphere in the 100--200 MHz band have largely been limited by the accuracy of these beam models.

We mitigate this problem by selecting sources within a narrow declination range, so that source tracks represent
nearly identical cuts through the beam response pattern.  Using this fact, the relative amplitudes of sources
can be deduced with minimal reliance on an accurate beam model.  In our analysis, a prior beam model is only
necessary for extrapolating across a narrow declination range (we estimate the errors associated with this 
extrapolation in \S\ref{sec:flux_scale}), and for approximating optimal inverse-variance weighting when averaging
over source tracks to determine a single spectrum.

This approach addresses several sources of error that were identified in \citet{Jacobs:2013p9908} 
as originating from uncertainty in the primary beam model.
  First, under most imaging schemes, each source flux is measured at just a few 
points in the primary beam. As errors tend to vary across the beam it is often
 difficult to decouple flux uncertainty from beam uncertainty.
 Using the drift scan-beamforming technique, each 
source is measured thousands of times at a variety of different primary beam values to provide a complete sample of the
flux variance due to primary beam variation.  Second, the flux calibration was found to vary over the 
sky due to the dual uncertainties of primary beam response and prior catalog. To reduce our exposure to flux calibration
variation, we limit our observations to sources passing within 5\arcdeg{}  of pointings which we can directly calibrate
to a single reference source.
 A third limitation, identified in \citet{Williams:2012p8768},
was the increase of uncertainty towards the edge of the beam. To minimize our sensitivity to both beam and noise
uncertainty, we weight the source track by an additional factor of the primary beam model.  

The remainder of the analysis in this paper relates to calibrating the relative source amplitudes that we
establish to an absolute flux scale as a function of frequency.  Given the shortcomings of 
existing in-band measurements of sources
in the southern hemisphere, we must bootstrap this flux scale from an ensemble of calibrator sources that
exhibit simple power-law spectra.
As described in \S\ref{sec:flux_scale},
this proceeds by first using a single calibrator to set the relative amplitudes of spectral channels, and
then using several sources, extrapolating across a large set of measurements 
between 70 MHz and 2 GHz, in a Markov-Chain Monte Carlo simulation to establish an absolute flux scale
for the measured spectra.

\subsection{Source Selection} 
As PAPER is a drift scan instrument, each declination describes a distinct path through the primary beam. The flux time series of a beamform
 provides a detailed sample of the primary beam relative to the peak of the trace, however,
a model is still required to calibrate between declinations. Below we argue that by averaging over long tracks, we minimize our 
susceptibility to localized primary beam error and find that limiting our source selection range to within 
a declination range of 5\arcdeg{} around a calibrator is a good compromise. In \S\ref{fig:gain},
we estimate the error resulting from this limited scope application of the primary beam model.

\begin{figure*}
\centering
\includegraphics[width=6in]{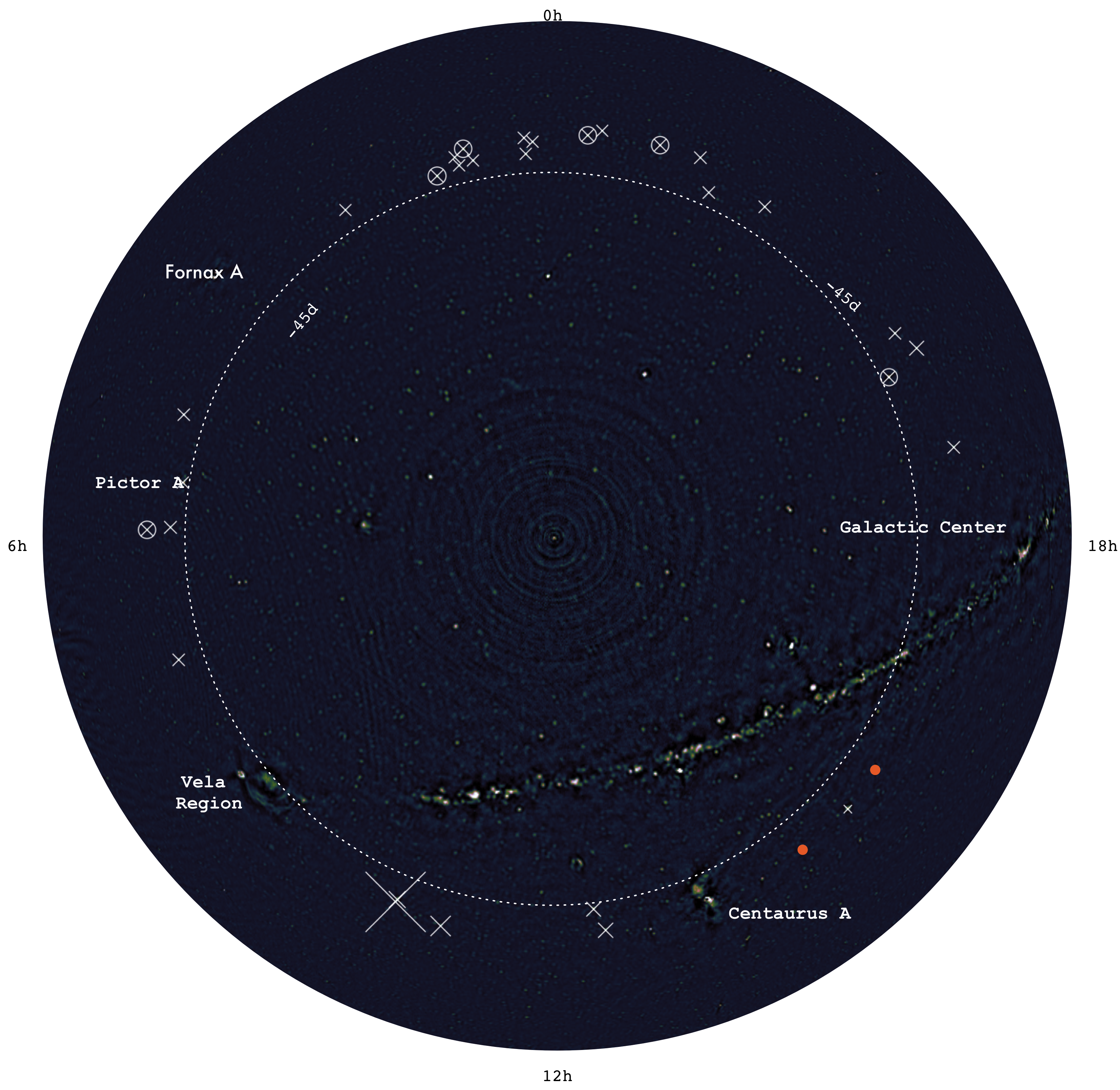}
\caption{
A PAPER map of the radio sky centered on the south pole showing relative positions of targeted sources. The map was made using the July half of the
data used in this catalog using the procedure described in \citet{Jacobs:2011p8438} to produce a lightly CLEANed map. The circular artifact around the south Celestial pole is due to residual instrumental cross-talk and appears to be highly localized below declinations of 75\arcdeg{} where PAPER's sensitivity is small.  `x''s mark measured source locations and have size
scaled by disagreement with prior data (inverse ``improvement index" from 
\S\ref{sec:results}). Smaller indicates better agreement with past data.  Orange dots indicate sources with no model overlap at 76\% in Figure \ref{fig:SI_contour_new}. 
\label{fig:error_map}
}
\end{figure*}

We choose sources from the Molonglo
Reference Catalog \cite[MRC]{Large:1981p7798} that are within 5\arcdeg{}  in
declination of a common calibrator (J2331-416), and more than 10\arcdeg{} from the galactic plane and
Centaurus A. We limit the flux extrapolated
from 408MHz to be greater than 10Jy, assuming a power law spectral index of
-1.\footnote{Most radio sources in this band have power law spectra $S(\nu) =
\left(\frac{\nu}{\nu_0}\right)^\alpha$ and a typical spectral index of $\alpha = -1$. } This selection contains 32 sources in a narrow
stripe that passes through the majority of the southern EoR fields. See Figure \ref{fig:error_map} 
for a map of the sources relative to the galactic plane and other structure as mapped by PAPER using the 
same observations presented here
and Table \ref{tab:data} for a complete listing of names and positions.

\subsection{Analysis Overview}
\label{sec:data_overview}
The spectra reported here are measured by beam-forming --- phasing the visibilities to
the target location and summing over baselines to produce a spectral time series of ``perceived'' flux $S_p$.
We define a perceived flux as the true source flux density, $S$, illuminated by the true beam pattern $B$:
\begin{equation}
S_p=B(t)S,
\end{equation}
where $B$ is a function of time because the observations are taken as a drift
scan.
The desired source spectrum is then isolated
from neighboring sources by delay filtering \citep{Parsons:2009p7859}.
We then average the spectrum in the time domain,
weighting by a model of PAPER's primary beam response ($B_M$); as the perceived flux is already weighted once by the true beam $B$, this weighted sum is an effective approximation of inverse-variance weighting \citep{Pober:2012p8800}.
Mathematically, our estimated source flux density is given by: 
\begin{align}
S_\textrm{est} &= \frac{\sum_t B(t)_M S_p}{\sum_t B(t)^2_M} \nonumber \\ 
&= \frac{\sum_t B(t)_M B(t)}{\sum_t B(t)^2_M} S  \nonumber \\
&\equiv g(\delta) S  \label{eq:g},
\end{align}
where the subscript $M$ indicates a model beam, and the $S$ without subscript is the true flux density of the source.
The net time integrated, weighted beam response $g$ is purely a function of declination $\delta$.  The rate at which this factor
changes with declination defines the declination range over which our calibrator source can be used to set the gain scale of the observations.
To simulate this affect we multiplied a measurement of the primary beam \citep{Pober:2012p8800}
by electromagnetic simulations \citep{Parsons:2010p6757} and find that $g$ 
diverges slowly with declination, with a maximum of 10\% over 5\arcdeg{} of declination range.
Thus, to first order,
we can use our calibrator source to calibrate sources within 5\arcdeg{} of declination and remove most of the variation with declination.
We estimate this spectral dependent $g$ at a reference declination of 41\arcdeg{}36\arcmin{} using (J2331-416), fitting a spectral index model to the
 catalog data points below 2GHz. This levels the bandpass due to spectral dependence of the primary beam, and allows us to place our measured source flux
estimates
on approximately the same flux scale.  
 Finally we average the spectra from a resolution of 400kHz to 10MHz bins, adding the individual errors in quadrature
 for our uncertainty.

At this point we have a set of spectra that have been calibrated relative to each other to the best of our ability. Now 
they must be put on a global flux scale. Meanwhile, our error estimate does not include an estimate of the 
flux calibration error.  To address both points we fit for a global flux scale that incorporates the best available
catalog data. These catalogs
have all been set, as best as possible, to the \citet{Baars:1977p9678} scale to which all other points are tied. Using a Bayesian analysis we compute the 
variation in the flux scale implied by comparing many sources to prior catalog models. By including sources from across
the RA/Dec range, the variation in this flux scale fit estimates the overall uncertainty in flux resulting from primary beam 
or prior catalog uncertainty.  This folds in the error due to extrapolation beyond the frequency range of the Baars scale which is
technically limited to 300MHz and above.

 \section{Data Reduction}
 \label{sec:Observations}

\begin{figure}\centering
\includegraphics[width=0.85\columnwidth]{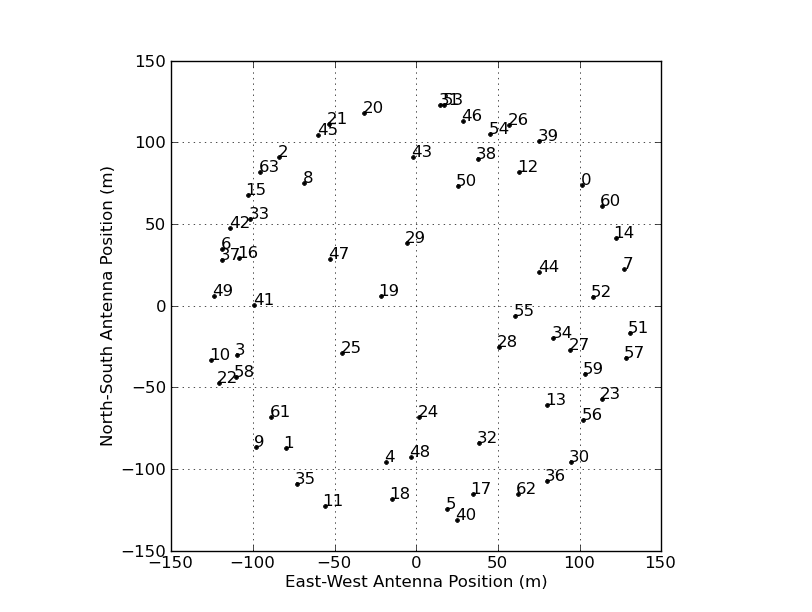}
\caption{Antenna positions in the 64-antenna, minimum-redundancy PAPER array configuration at the Karoo Radio Observatory site in South Africa used for these observations.
Data was captured during July and October of 2011 on a single 'x' polarization over a band between 120 and 180 MHz.
}\label{fig:antpos}
\end{figure}

 Measurements are derived from observations using two nights on the east-west dipole arms of
64 PAPER antennas deployed in a minimum-redundancy imaging configuration
(see Figure \ref{fig:antpos})
at the Karoo Radio Observatory site in South Africa.
A100-200MHz band was correlated
with 2048 frequency channels and integrated for 10.7 seconds before
visibilities were stored.  Observations included here on  July 4, 2011, running
from JD2455748.17 to JD2455748.72. and October 17, 2011 running from
JD2455852.2 to JD2455852.6.  In both seasons two antennas were omitted owing to
malfunctioning signal path. This 20 hour long combined dataset provides
observations provides near complete hour angle coverage for the entire 24 hours of
Right Ascension. The July portion of the dataset comes from the same observing
campaign described by \cite{Pober:2013p9942} and \cite{Stefan:2013p9926}.

\subsection{Data Compression}

In pre-processing, we use delay/delay-rate (DDR) filters
\citep{Parsons:2009p7859} to identify radio-frequency interference (RFI)
events, and as part of a data compression technique that reduces data volume by
over a factor of 40.  A more detailed description may be found in Appendix A of
\cite{Parsons:2013p9876}, which uses the same compression method on PAPER power
spectrum observations. Here we summarize the process.

First, we remove known RFI transmission bands and analog filter edges, and then
flag outliers in the frequency and time derivative at 6$\sigma$ to remove RFI events.  Next, we suppress sky-like
emission by applying a DDR filter to remove delays and delay-rates within the
horizon limit of a 300-m baseline (the maximum length of any PAPER baseline).
We derive a second set of RFI flags by masking 4$\sigma$ outliers in these
residuals and apply these flags back to the unfiltered data.  Finally, we
compress the data by applying a DDR filter preserving emission within the
horizon limit of a 300-m baseline, deconvolving to suppress flagging artifacts,
and down-sampling the result in time/frequency domain to the critical Nyquist
rate of the DDR filter.  The result is that the 2048 original frequency
channels become 203, and the 60 original time samples per 10-minute file become
14.

\subsection{Per-Antenna Gain Calibration}

Small time- and antenna-dependent gain variations time introduce significant systematic effects that degrade 
instrument performance. Temporal variations are dominated by changes in ambient temperature. 
We use measurements of ambient temperature versus time on the balun amplifier of a fiducial antenna and near
receiver amplifiers to divide out the predicted gain variation versus temperature that these amplifiers
are known to exhibit.  Inspecting the auto correlations we find that the net antenna temperature gain coefficient (-0.042 dB/C) found in previous laboratory \citep{Parashare:2011p9872} and early field 
measurements  \citep{Pober:2012p8800} did not appear to remove all temperature variation.  To
revise the temperature coefficient we use the October observations, when the temperatures remained nearly constant, as a reference.

\begin{figure}
\includegraphics[width=0.9\columnwidth]{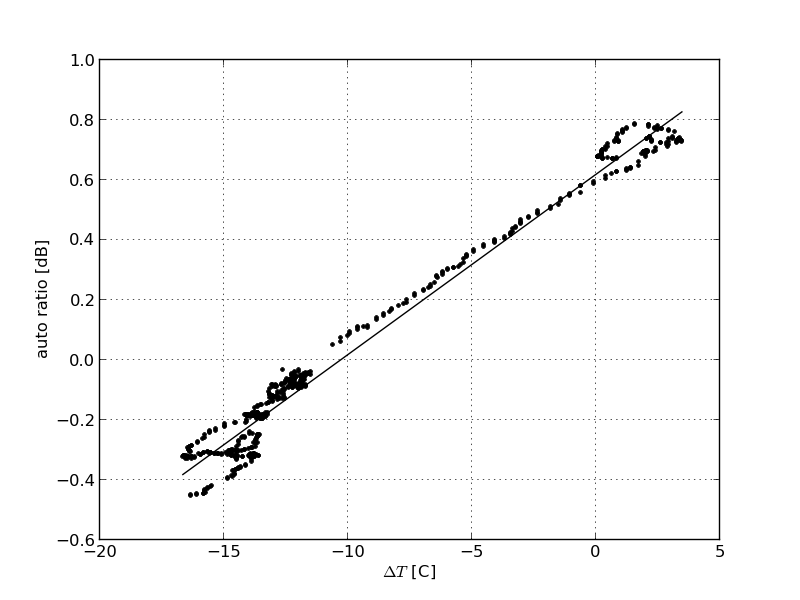}
\caption{Calibrating the temperature coefficient  between the July and October nights.  Here we plot the ratio of auto-correlations (averaged over frequency and antenna) against the temperature difference between the two nights. The very linear middle region has a slope of  -0.055 dB/C while the full, more complex, range is fit by -0.06dB/C (black line).  We compromise by choosing -0.058dB/C   
which was found to significantly improve the agreement (see Figure \ref{fig:auto_compare}).
\label{fig:autos_tempcorr}}
\end{figure}

\begin{figure}
\includegraphics[width=0.99\columnwidth]{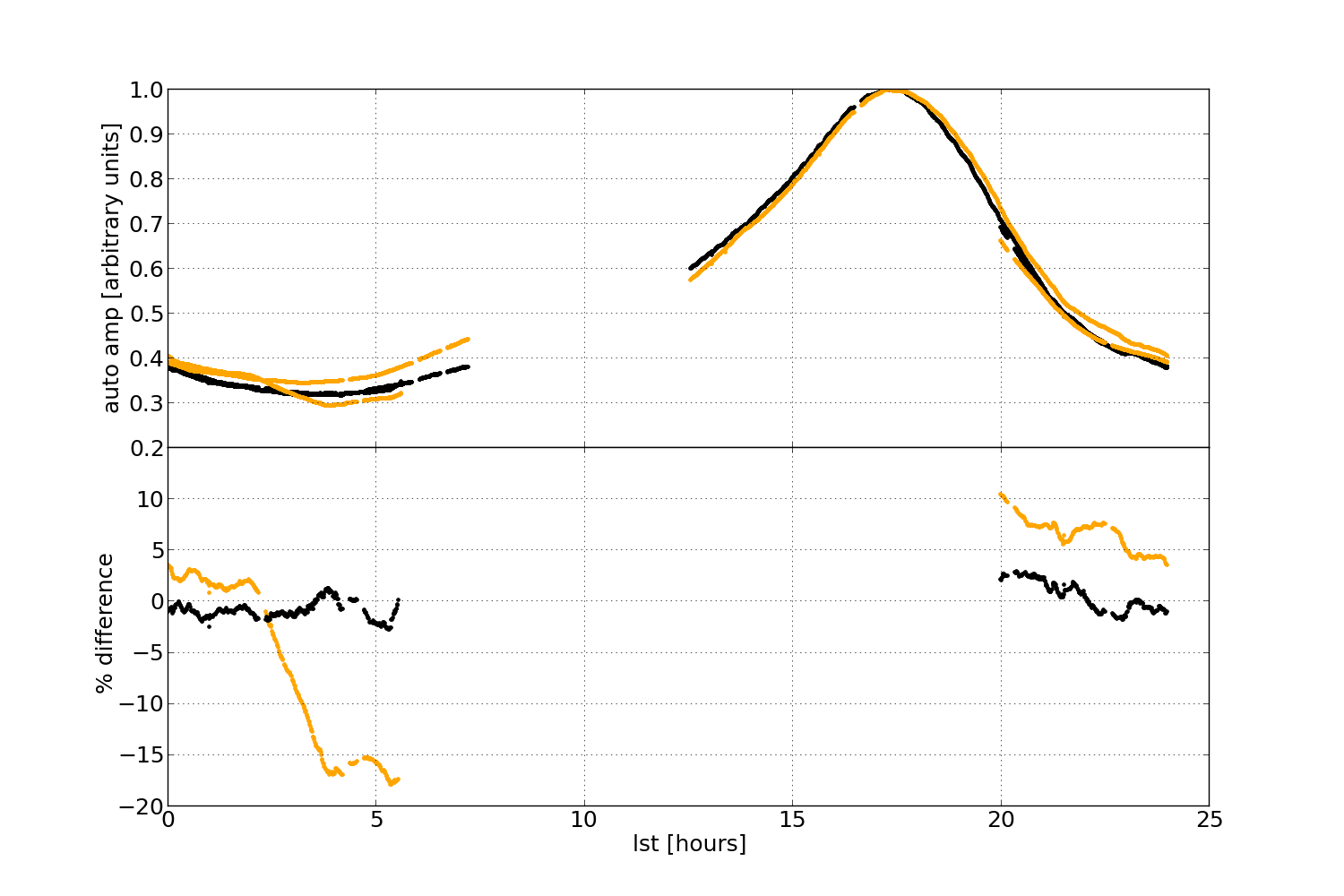}
\caption{Gain differences between the July and October nights are dominated by temperature variation. Here we see the effect on 
the average auto correlation before ({\color{orange} orange}) and after (black) application of the temperature dependent gain. \label{fig:auto_compare}}.   Note that the span of the data in the bottom fractional difference plot is limited to points occurring in both nights and does not cover the full range seen in the top.
\end{figure}

Comparing the average ratio between the July and October autocorrelations to their temperature difference (Figure \ref{fig:autos_tempcorr}) we find that the two are highly correlated  with 
a best fit temperature coefficient of (-0.058 dB/C). As we see in Figure \ref{fig:auto_compare}, this removes a significant amount of disagreement between the two nights, 
with the peak difference decreasing from 18\% to 3\%.

Matching of relative gains and phases between antenna was done by fitting a per-antenna complex gain to portions
of data which have easily modeled sky. 
This calibration is only relative between antennae, bandpass and absolute flux calibration
comes after time and frequency averaging in sections \ref{sec:Calibration} and \ref{sec:flux_scale}.   
The antenna delays and amplitudes were found by fitting a point source visibility
model to Centaurus A, Pictor A, and Fornax A.  Each source is imperfectly
modeled by a single point-source, but the solution differences are minimized by
averaging over the three independent solutions. These same calibration
solutions have been successfully applied in \citet{Pober:2013p9942} and \citet{Stefan:2013p9926}.
for power-spectrum analysis of foregrounds and imaging of Centaurus A, respectively.  

The October per-antenna gains were found by computing the average ratio of the auto correlations
against the July data. Delays were held constant as in \citet{Jacobs:2011p8438}.


\subsection{Beamforming}
\label{sec:beamforming}
Spectral time-series are computed by beam-forming to the selected sky
locations. A beam is formed by phasing baselines to the desired location and
summing over baselines longer than 20 wavelengths.

\begin{figure}
\includegraphics[width=0.9\columnwidth]{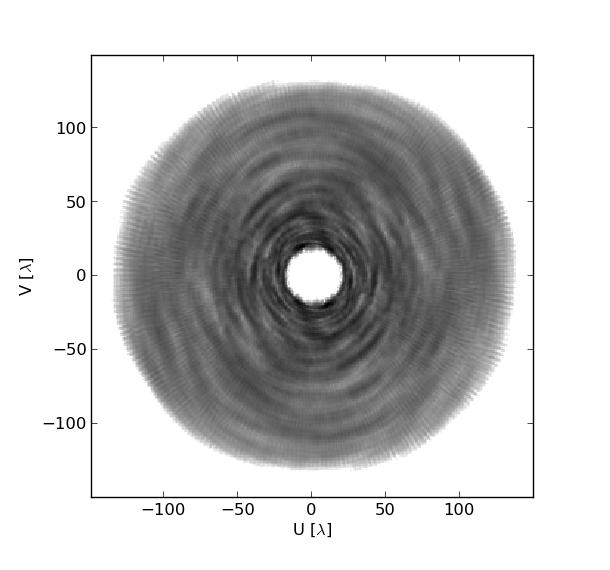}
\caption{The effective uv coverage of a 10MHz-wide spectrum bin, showing the relative contributions of each baseline when beam-forming on Pictor over a 9.6 hour synthesis in the October observing session. \label{fig:uv_coverage}}
Not shown here are the weights interpolated from the Perley et al 330MHz map (Fig. \ref{fig:pic_perley}) used to recover the total integrated flux of Pic  which represent a few \% change on the longest baselines.
\end{figure}

 These complex spectra are then fringe rate filtered to remove spectrally
smooth sources that deviate more than $\pm$0.1mHz from the fringe rate of the source in question
 \citep{Parsons:2009p7859} (cf the LOFAR
de-mixing approach \citep{Offringa:2012p9691})  producing a time dependent
spectrum with minimal side lobes. This is then averaged in time,  weighting by a model of the primary
beam as discussed above.
The result is equivalent to
a very long earth rotation synthesis image with a single image pixel. The weighted
contributions from each baseline (the uv coverage) are shown in Figure \ref{fig:uv_coverage}.  When
combined with the filtering it is a robust and simple method for measuring
spectra of unresolved sources. 

\subsection{Compensation for Resolution Effects in Pictor A}

\begin{figure}
\includegraphics[width=0.96\columnwidth]{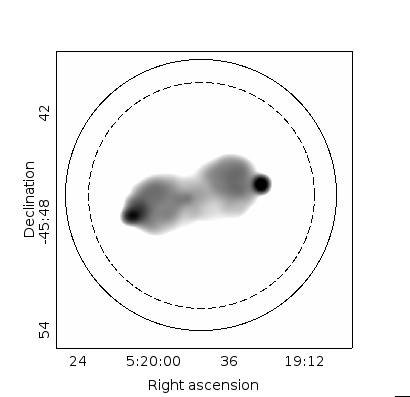}
\caption{
Pictor A imaged at 330MHz by the VLA \citep{Perley:1997p9312}. Black circle
indicates PAPER resolution at the 150MHz, dashed, the resolution at 185MHz.  To
correct for the small residual structure in the PAPER measurements we resampled 
(via spline interpolation in the Fourier plane)
the Perley image to the PAPER resolution and estimated the relative contribution
 to each PAPER baseline.  At the highest frequency this
led to a 3\% correction, at the lowest frequency, 0.6\%.
\label{fig:pic_perley}}
\end{figure}

\begin{figure*}[htbp]
\includegraphics[width=0.49\textwidth]{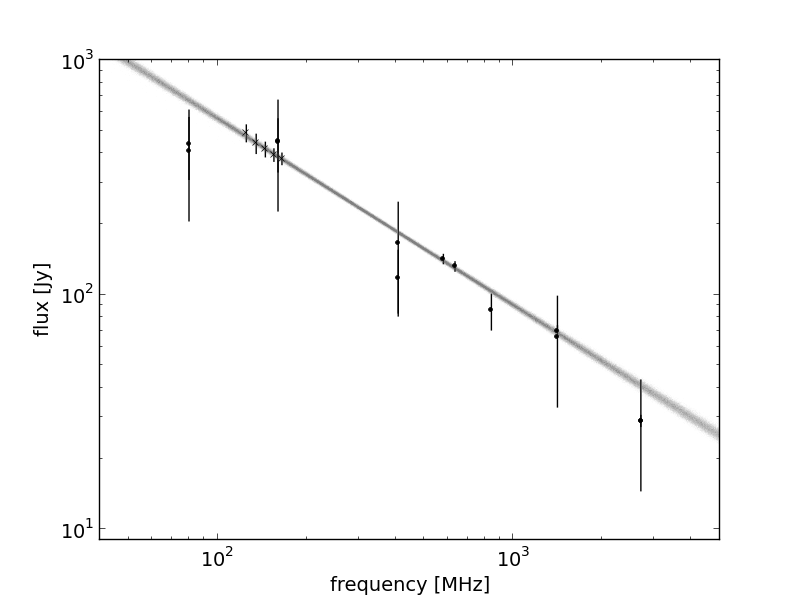}
\includegraphics[width=0.49\textwidth]{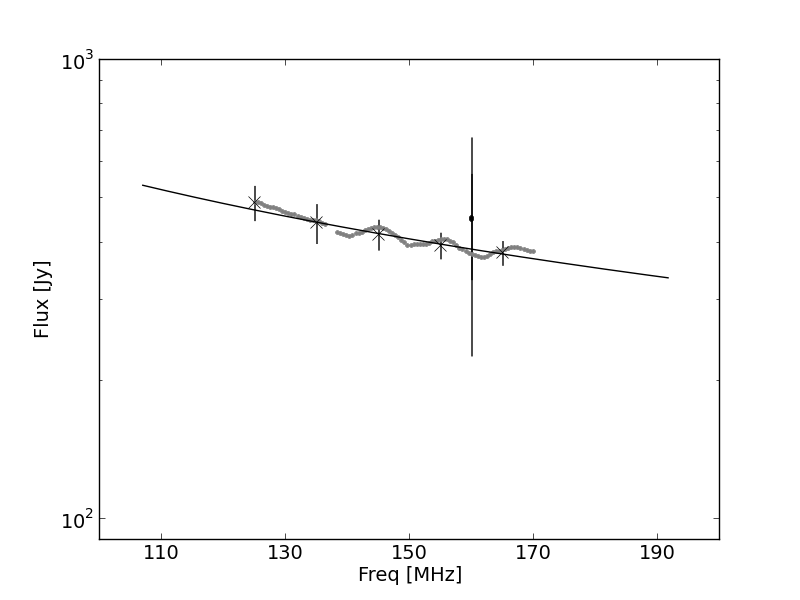}
\caption{
Using new PAPER results to constrain the spectrum of Pictor A.  On the left,
the PAPER Pictor A spectrum with 2$\sigma$ error bars (x) and  MCMC fits (grey
band). On the right, the spectrum zoomed to highlight the PAPER data
 allows comparison between the 10MHz averaged points with the
403kHz points (grey). The oscillations in the PAPER spectrum are residual side
lobe contamination from other sources and are well represented by the error
bars.  The most reliable prior measurements (dots) are those by
\cite{Wills:1975p9314} at 580MHz and 635MHz \citep{Perley:1997p9312}. The rest
are from with Culgoora \citep{Slee:1995p7541} or Parkes
\citep{Otrupcek:1991p8780} and have been set to the (extrapolated)
\citet{Baars:1977p9678} scale by  \cite{Kuehr:1981p9628}. 
} \label{fig:pic_spectrum}
\end{figure*}

The beam-forming method assumes the target is a point source. Our primary
target, Pictor A, is slightly resolved by PAPER, and merits closer attention.
Pictor A is a double-lobed radio galaxy with a main lobe separation of
7\arcmin{}. As we see in Figure \ref{fig:pic_perley}, it is nearly unresolved by
PAPER's 15\arcmin{}  synthesized beam. However, given the high SNR of the
observations, we see a 20\% drop in flux on the longest ($\sim$300m) baselines. 
We account for this by weighting baselines
in the beamform step according to a model of structure observed by
\citet{Perley:1997p9312} at 330MHz, which they found to be consistent with
their more limited 74MHz images as well as the detailed high frequency maps.
The normalized image is Fourier transformed and sampled at the desired uv
spacing by spline interpolation. These samples are used to weight each baseline
contribution in the baseline sum. The result is an estimate of the total
integrated flux for Pictor A. Where the resolution is highest, at the top of
the band, the net correction is 3\%. At the bottom of the band, where the
resolution size has grown to 19\arcmin{}, the correction is only 0.6\%. The
resulting spectrum is shown in Figure \ref{fig:pic_spectrum}a.

\subsection{Bandpass Calibration}
\label{sec:Calibration}

The resulting set of spectra were then calibrated to a model of J2331-416 
\footnote{Here we set the spectrum to $S_{150}=33$Jy, $\alpha=-0.76$, derived by a linear least squares fit to
catalog measurements below 1GHz \citep{Slee:1995p7541,Kuehr:1981p9628,Large:1981p7798,Burgess:2006p7814}.  Further calibration on top of this somewhat arbitrary calibration model
is explored in detail below. }
 to remove 
the residual bandpass $g$ from equation \ref{eq:g} caused by the net effect of the primary beam as discussed in section \ref{sec:approach}. 
This source
was selected for its relatively high brightness, spectral smoothness and large quantity
 of available catalog data. Each source track has a slightly different sample profile
 resulting from different amounts of flagged data at each time and channel. To account
 for these differences in the bandpass calibration we build a set of calibration tracks 
 that match the tracks for each source. Calibration of each source then proceeds with
 an optimally matched calibration spectrum.

\subsection{Fitting a Power-Law Model to Spectra}
\label{sec:mcmc}

There is a variety of prior data at multiple wavelengths to which we want to calibrate
and then compare our measurements.  Our method for doing both of these steps
is to assume a basic spectral model relating the different catalog data points and
fit for spectral model and gain parameters. 

We estimate spectral model parameters and flux calibration in a
Bayesian way by calculating and marginalizing the posterior probability of the
catalog and new PAPER data.   This method
offers improved repeatability by specifying a single objective function that
represents the quality of the model fit and naturally defines the errors on the
parameters \citep{Hogg:2010p8759}.   See  \citet{Mackay:2003p9717}  or
\citet{Sivia:2006p9736} for more on Bayesian analysis methods and an excellent
astrophysical example by \cite{Press:1997p9783}. In brief, measurement errors
are related to parameter model errors via the likelihood, which we can calculate, 
and the posterior, which we cannot.  However, the posterior is theoretically
well sampled by a Markov Chain Monte Carlo (MCMC) sampler, which selects parameter
values at random, computes the likelihood of the model given the data and noise model,
and accepts or rejects the step based on an outside decision factor unrelated to the data.

Here we use the \emph{emcee} sampler by \citet{Mackay:2003p9717} et al to generate chains of
parameter values. The best fit model is the median of all the sampled
parameter sets, while the volume containing a well defined fraction of samples
sets the confidence limits.  The oft quoted ``1$\sigma$" probability level
corresponding to a gaussian distribution contains 65\% of the samples. In
practice the contours are not gaussian. Here we choose a slightly more
conservative probability level of 76\%

To model the relationship between different wavelengths we assume a single spectral index
which is the prevailing spectral energy distribution at low frequencies, 
though curvature or other
deviation from a power-law is not uncommon.  

When fitting models to the full catalog, we use the \citet{Vollmer:2010p6422}
catalog which has been optimally cross-matched at the expense of excluding more data points.
Meanwhile, the gain fit used a small sample of
sources with spectra which meet our calibration criteria: more precision data available,
brighter than most sources in the sample and far from any possibly confusing areas of
the sky. 
By limiting ourselves to a small number of sources we are able to go include catalog points by hand,
to avoid making the error of falsely including erroneously matched 
data points.

\subsection{Approximating the Absolute Flux Scale}
\label{sec:flux_scale}

At the output of the beam forming and bandpass calibration step, the flux scale
 is tied to a model fit of the catalog values of
J2331-416. The accuracy of this fit, and the implied uncertainty in the flux
scale, limits the accuracy of the PAPER measurements.  To refine the flux
scale and estimate our flux scale uncertainty, we bootstrap a single global flux scale 
correction factor using 6
sources selected for their brightness, spectral linearity and data
availability\footnote{Calibration sources:
2250-412,2331-416,1932-464,0103-453,0547-408,0043-424}. To build a more complete
spectral model we go beyond the data found in \citet{Vollmer:2010p6422} to include all spectral
measurements below 2GHz found in the NED database \footnote{\url{ned.ipac.caltech.edu}: Accessed 1 April 2013}.  These additional
 catalog measurements are primarily by Parkes, and Molonglo  \citep{Kuehr:1981p9628} with the best precision coming from the
Wills fluxes at 538 and 634 MHz \citep{Wills:1975p9314}. Where error bars are not
given, we assume uncertainty of 25\%. 

\begin{figure}
\centering
\includegraphics[width=0.9\columnwidth]{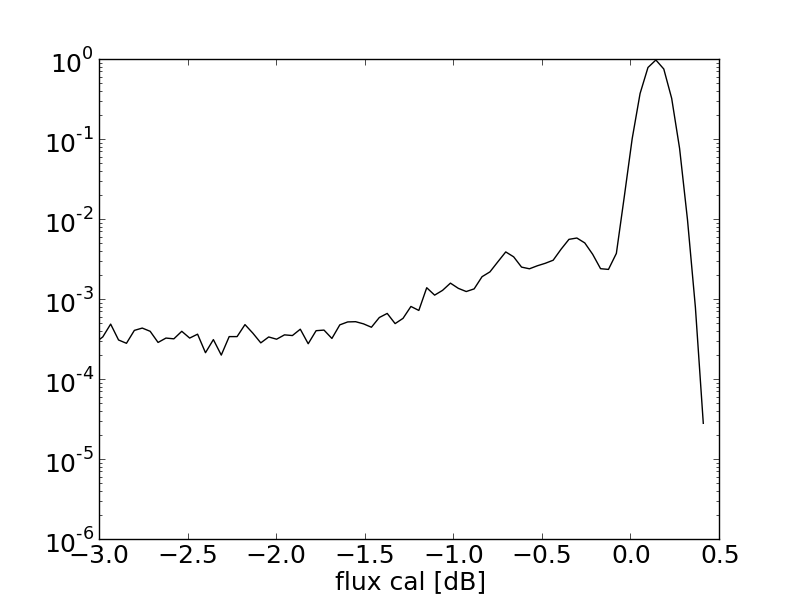}
\caption{
The marginalized posterior of the PAPER flux scale factor relative to 2331-416. Given a joint fit to all the
 NED database listings for
2250-412,2331-416,1932-464,0103-453,0547-408,0043-424 below 2GHz and the corresponding PAPER values. The peak of the distribution sets the overall flux scale,
while the 76\% width is an additional
 fractional error to be added quadratically to the per-source uncertainty.  
By averaging over many sources and and Baars calibrated data points, this uncertainty encapsulates error due to both primary beam model inaccuracy 
as well as the average flux scale uncertainty due to extrapolating many absolutely calibrated measurements.
\label{fig:gain}}
\end{figure}

Using the MCMC method we fit spectral index models to the catalog fluxes
 simultaneously with a global PAPER flux scale factor using  an MCMC chain to calculate the log likelihood

\begin{align}
\log\mathcal{L}_s = &\sum_{\nu}\frac{ \left[S_\textrm{cat}^{\nu}  - S_{150}  \left(\frac{\nu}{150}\right)^\alpha\right]^2}{2(\Delta S_\textrm{cat}^\nu)^2} \\
&+\frac{ \left[g S_\PAPER^{\nu}  - S_{150}\left(\frac{\nu}{150}\right)^\alpha\right]^2}{2(\Delta S_\PAPER^\nu)^2} \nonumber \\
\log\mathcal{L} =& \sum_{s} \log\mathcal{L}_s
\end{align}

which samples the posterior probability of PAPER data ($S_\PAPER^{\nu}$) and
catalog values ($S_\textrm{cat}^{\nu}$) given a spectral index model for each source and
a global flux scale factor ($g$).
Marginalizing over the fitted flux scales, we find the resulting flux scale distribution function which is shown in Figure
\ref{fig:gain}. The 76\% confidence limit on this flux scale, relative to the
J2331-413 calibration, is  $+0.11 ^{+0.05}_{-0.08}$ dB, or a 1.54\%
multiplicative error on every calibrated PAPER measurement.  This flux scale is applied to the PAPER spectra
with the errors added in quadrature.

\begin{figure*}[htbp]
\begin{center}
\includegraphics[width=0.9\textwidth]{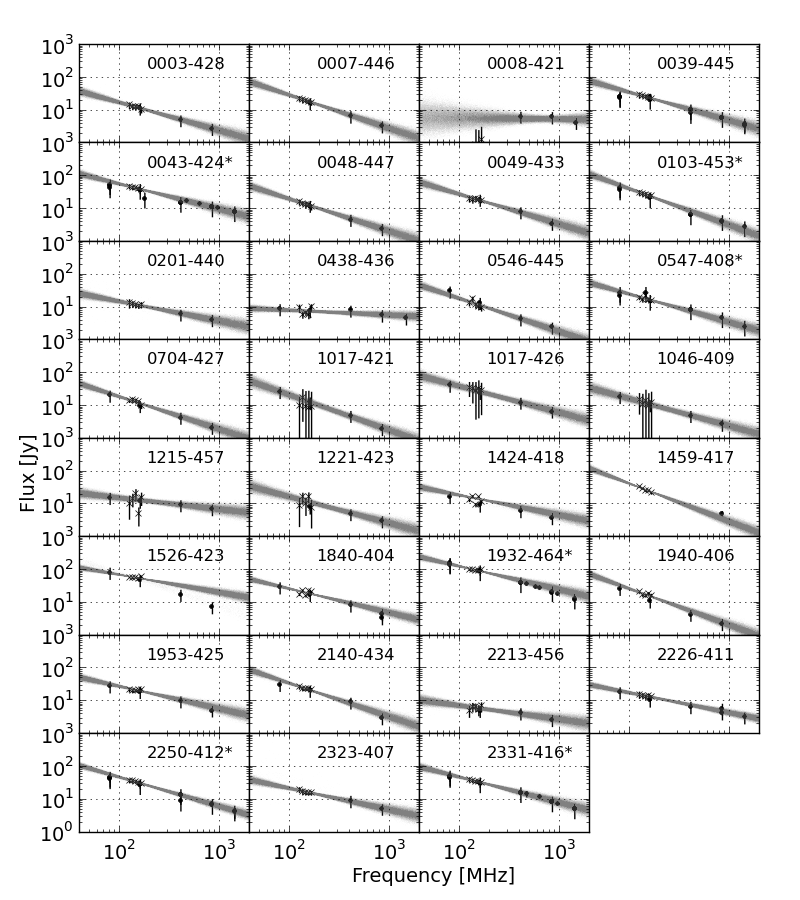}
\end{center}
\caption{
PAPER spectra of 31 sources compared against existing data out of
\cite{Vollmer:2010p6422} between 40MHz and 2GHz and a sampling
of spectra fits drawn from MCMC samples at greater than 76\% confidence. Sources used to bootstrap the
flux calibration are noted with a `*' and are shown with the additional calibration data
found in NED. Pictor A is shown separately in Figure \ref{fig:pic_spectrum}.
\label{fig:srcs1}}
\end{figure*}

These calibrated spectra are plotted in Figure \ref{fig:srcs1} and are listed in Table \ref{tab:data}.  The
resulting Pictor A spectrum ranges in precision from 3.8\% at 135MHz to 2.6\%
at 165MHz. About half of this error is due to uncertainty in the flux scale.

\subsection{Fitting Spectral Models}
\label{sec:fitting_models}

Finally we compare all of our  calibrated spectra to catalog measurements. Our method
will be to first establish a best fit model to existing data, then add the PAPER data
to the fit, and assess the degree to which the PAPER data is supported by prior measurements (accuracy), and the reverse,
the degree to which PAPER offers an improvement in our knowledge of the spectrum (precision).

 To establish a baseline model, we
fit a spectral model to prior catalog data from the spectrally and spatially
cross-matched meta-catalog by \citet{Vollmer:2010p6422} using the
 MCMC chain to sample the (log) likelihood which assumes Gaussian measurement errors

\begin{equation}
\logL = \sum_\nu{\frac{\left[S_\nu - S_{150}\left(\frac{\nu}{150}\right)^\alpha\right]^2}{\Delta S_\nu} }
\end{equation}

\begin{figure}[htbp]
\includegraphics[width=0.49\textwidth]{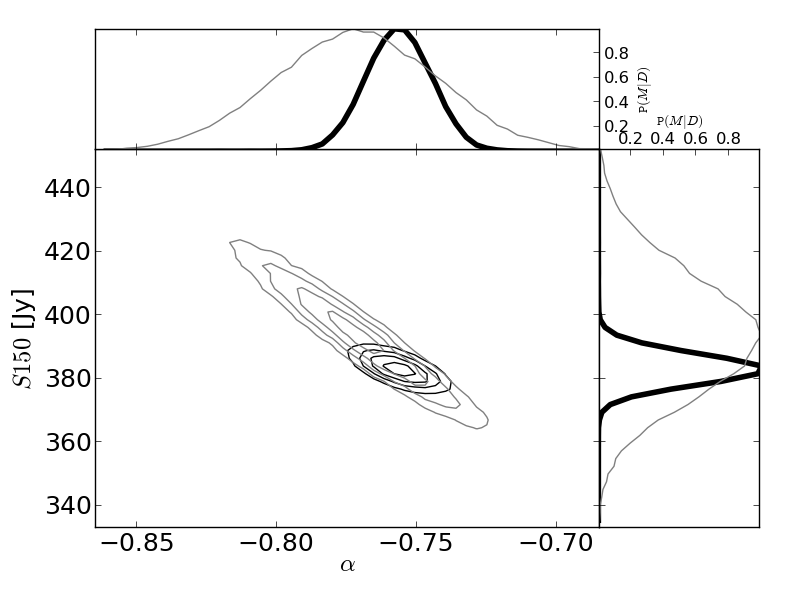}
\includegraphics[width=0.49\textwidth]{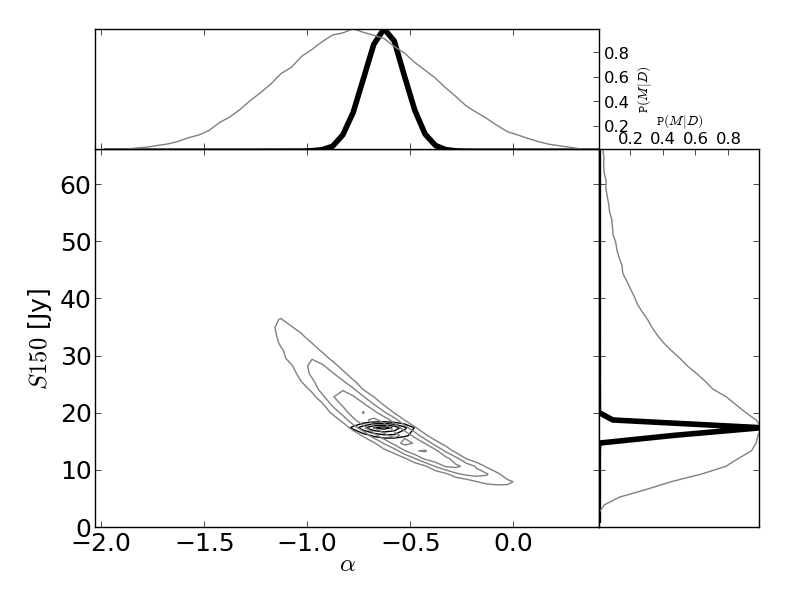}
\caption{
A detailed view of the posterior probability distribution as described in section \ref{sec:fitting_models}. On the left
Pictor A, on the right 2323-407. Confidence contours range from 0.2 to 0.8, curves top and right 
are the marginalized distributions. Grey plots the fit to catalog data, black for joint catalog and PAPER.
 The MCMC
chain reveals the distinctive ``banana'' shape of the posterior
due to correlation between the model parameters.
Compare with the contours for this source shown in Fig. \ref{fig:SI_contour_2}. Fit parameters listed in
Table \ref{tab:fits} are the locations where the marginalized curve crosses 0.76.
\label{fig:posteriors}
}
\end{figure}

As when fitting the flux scale, 
we estimate the confidence interval of the resulting parameters as the boundary
enclosing 76\% of the samples. A detailed view of the resulting posterior is shown in grey on Figure 
\ref{fig:posteriors}. Most of the posteriors are characterized by steep sided, symmetric 
distributions. While many appear to have linearly correlated parameters, several display
the classic banana shape associated with a non-linear correlation between parameters.
For the rest of the sources we focus on the 76\% confidence levels, plotting the 2D distributions
in Figures \ref{fig:SI_contour_1} - \ref{fig:SI_contour_bad}, and listing the marginalized 
values in Table \ref{tab:fits}.

The fit is then performed again with
both PAPER and catalog data, shown in black on the same Figures.

\section{Results and Discussion}
\label{sec:results}
The majority of the \Nsrc{} models fit using the new PAPER data were found to agree with
models fit to past data and all but two are suitable for comparison.  
0008-421 was not detected by previous PAPER observations and
exhibits flattening at higher frequencies, most likely due to synchrotron
self-absorption \citep{Jacobs:2011p8438}. Meanwhile, 1459-417 does not have enough 
measurements in the SPECFIND catalog on which to build a model for comparison. We exclude these 
two from further analysis.

Inspecting the remaining 30 confidence contours we find that, for the vast majority ($
90$\%), the PAPER measurements confirm the power law extrapolation from higher
frequencies. The remaining 10\% either A) have large PAPER error bars and
therefore provide no new information or B) do not agree with the spectral index
model.  To understand where most measurements fall on this spectrum we
numerically quantify the overall model improvement derived from the addition of
the PAPER data as 
\begin{fleqn}[0pt]
\begin{align*}
&\textrm{Improvement} =\\
&= \left(\textrm{Precision increase}\right) \left(\textrm{Catalog agreement}\right)\\
&= \left(\frac{1}{\textrm{Area}(\textrm{PAPER})} - \frac{1}{\textrm{Area}(\textrm{Cat})}\right) \left( \textrm{Area}(\PAPER \cap \textrm{Cat}) \right)
\end{align*}
\end{fleqn}
where the area is defined as the number of samples having posterior probability above 76\%.

We quantify the fit {\em precision increase} as the change in the contour
figure of merit, defined as the inverse area of the confidence contour.
Meanwhile, {\em catalog agreement} is the fraction of the PAPER confidence
interval that overlaps the catalog confidence interval.  Thus, for example, in
a PAPER fit that overlaps the catalog confidence contour by 41\% but increases
in precision (confidence area shrinks) by a factor of 3, the resulting {\em
improvement} will be 0.123.  This Improvement Index is included in the table
of fit parameters (Table \ref{tab:fits}) and is used as the size scale on the 
map in Figure \ref{fig:error_map}.

\begin{figure*}[htbp]
\begin{center}
\includegraphics[width=2in]{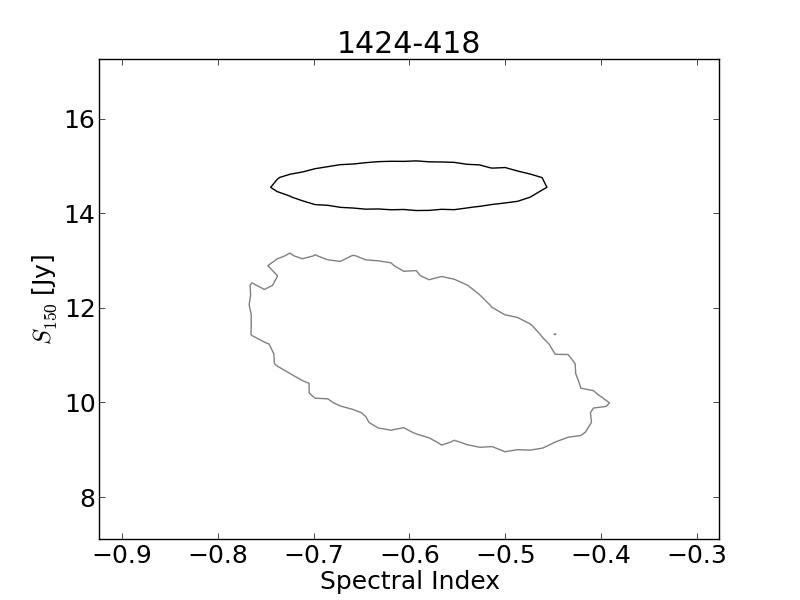} 
\includegraphics[width=2in]{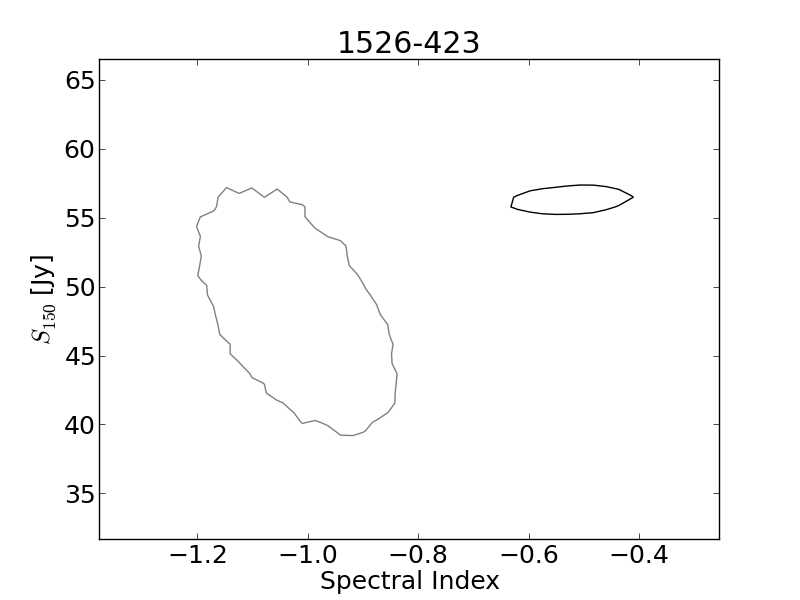} 

\end{center}
\caption{
These fits are somehow at odds with other measurements.  The 1526-42 spectrum is visibly curved (see Fig. \ref{fig:srcs1}), which explains the large disagreement in spectral index but not flux. 1424-418 is an optically polarized quasar with a flat and variable radio spectrum which might imply co-aligned jet viewing and high intrinsic variability. 
} \label{fig:SI_contour_new}
\end{figure*}

\begin{figure*}[htbp]
\begin{center}
\includegraphics[width=2in]{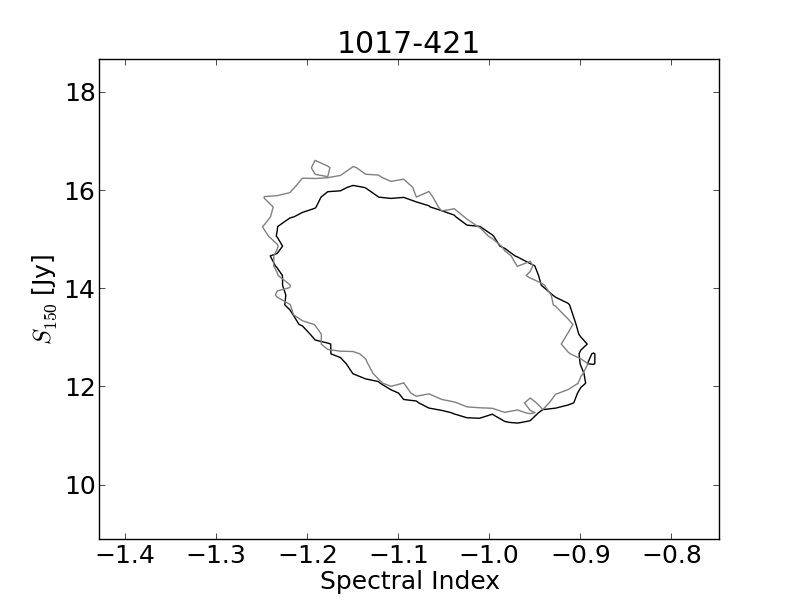} 
\end{center}
\caption{A source with a slightly negative improvement index (-0.0011).  Here the PAPER data represents a large fraction of the available data
but due to the large error bars slightly increases the uncertainty.
}\label{fig:SI_contour_bad}
\end{figure*}

In these sources, the improvement index ranges from a maximum of 7.84  to -0.001.
One source (1017-421) shows a slightly negative improvement, suggesting that PAPER data have
added to uncertainty (see Figure \ref{fig:SI_contour_bad}), while two sources
have exactly zero improvement, which indicates that the PAPER data have pulled
the fit far from the model preferred by the catalog data (see Fig.
\ref{fig:SI_contour_new} for details on these).  

\begin{figure*}[htbp]
\begin{center}
\includegraphics[width=2in]{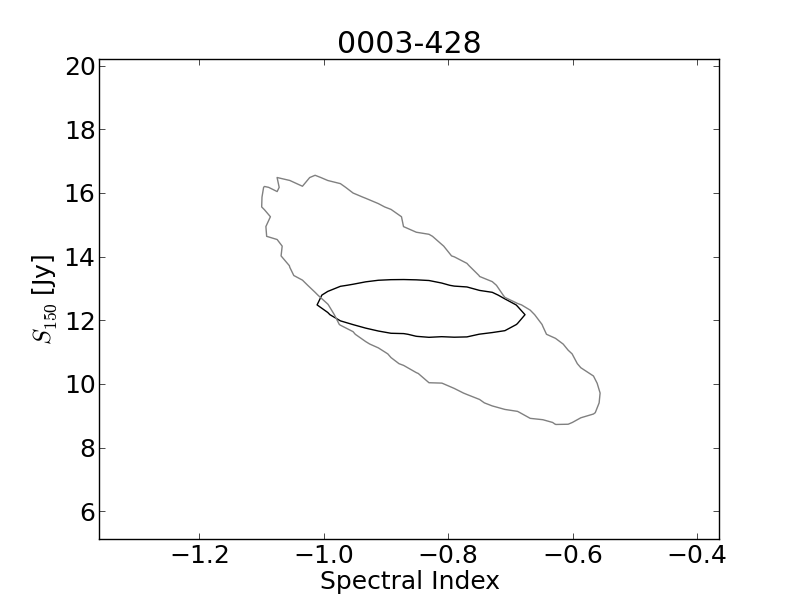} 
\includegraphics[width=2in]{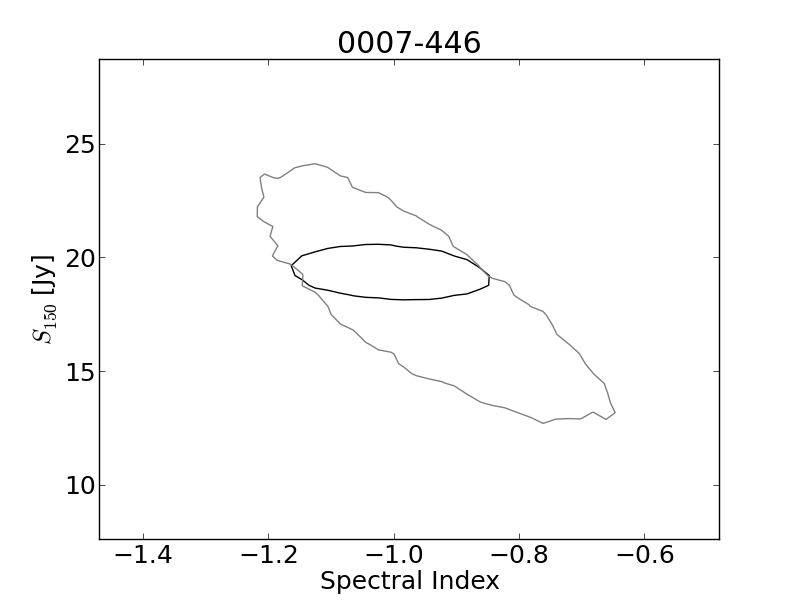} 
\includegraphics[width=2in]{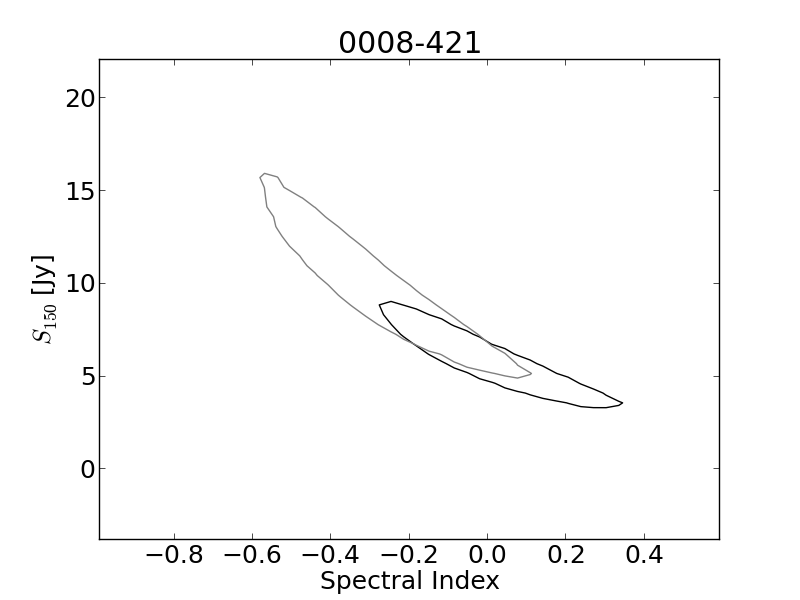} 
\includegraphics[width=2in]{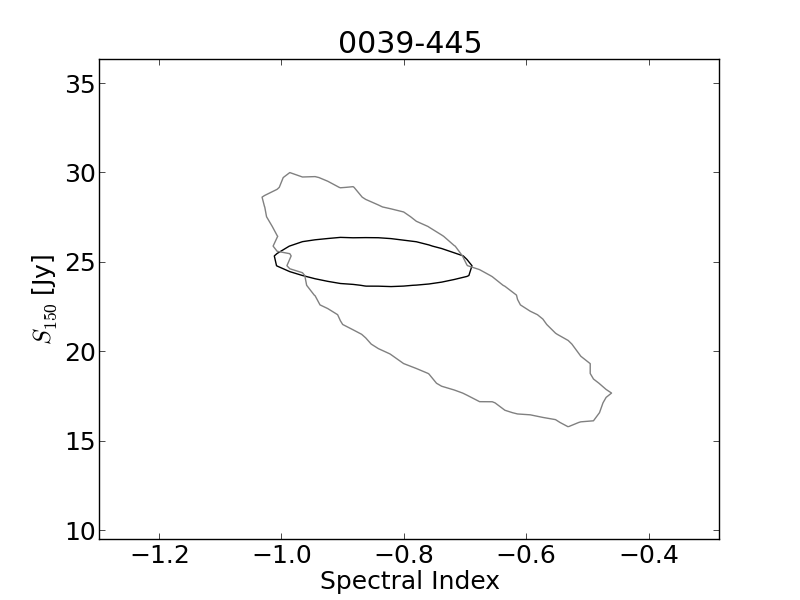} 
\includegraphics[width=2in]{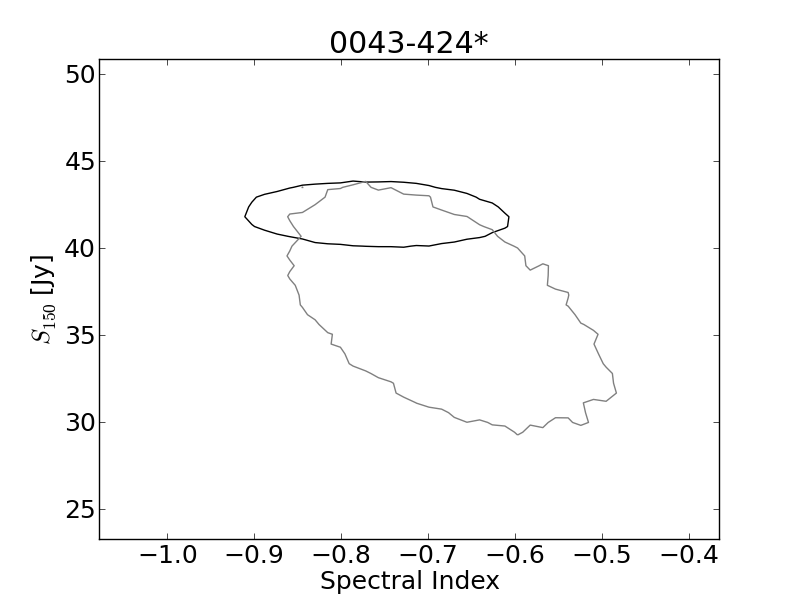} 
\includegraphics[width=2in]{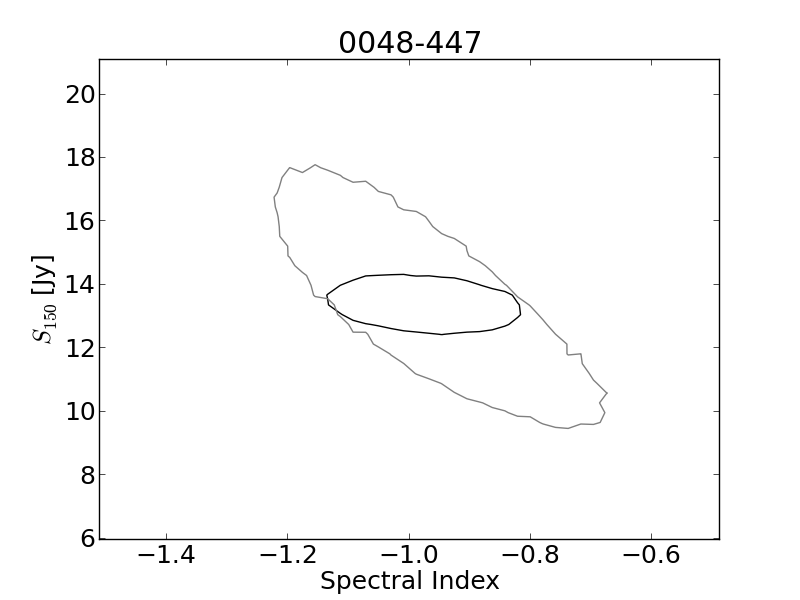} 
\includegraphics[width=2in]{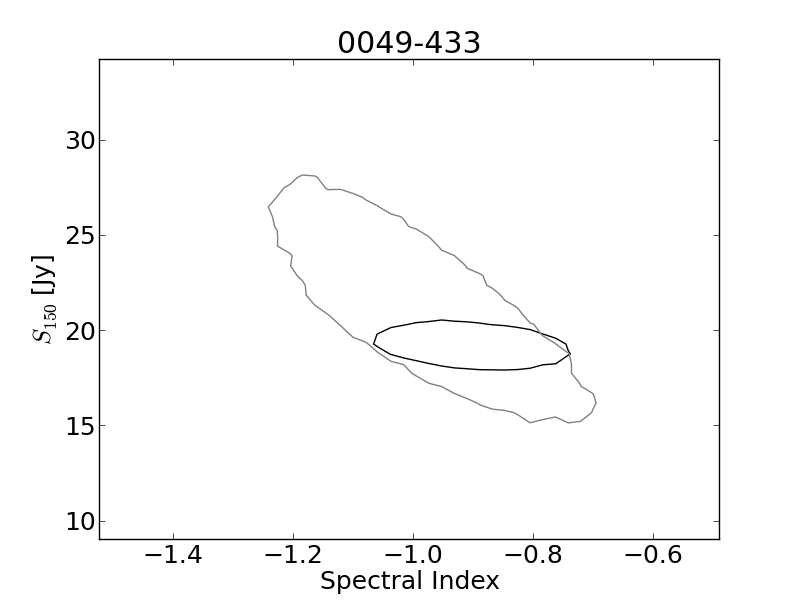} 
\includegraphics[width=2in]{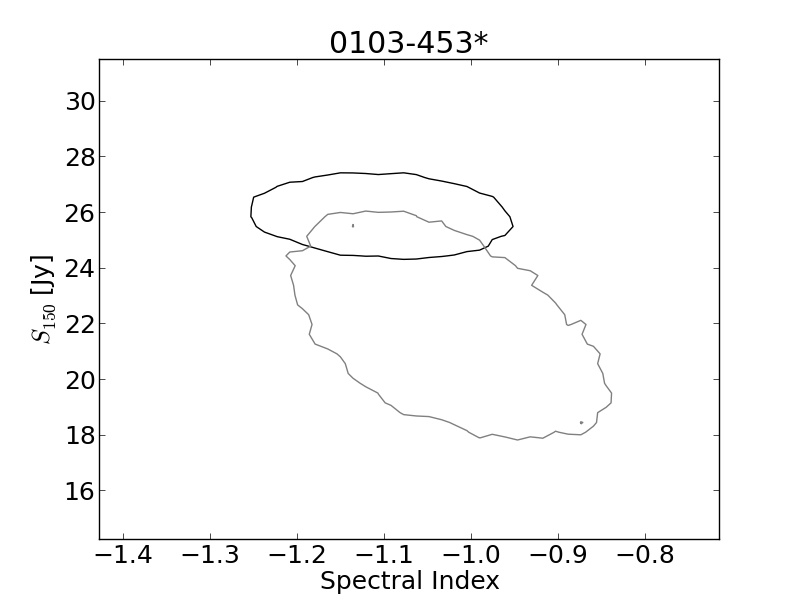} 
\includegraphics[width=2in]{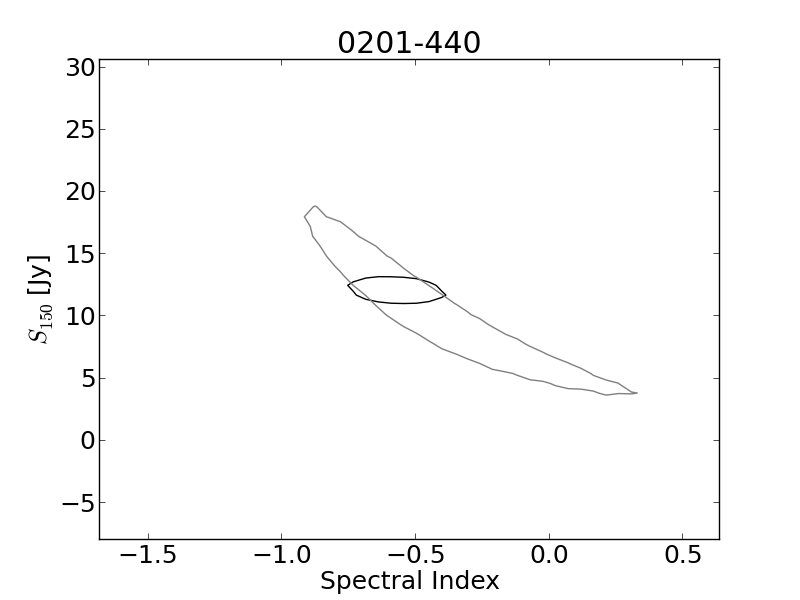} 
\includegraphics[width=2in]{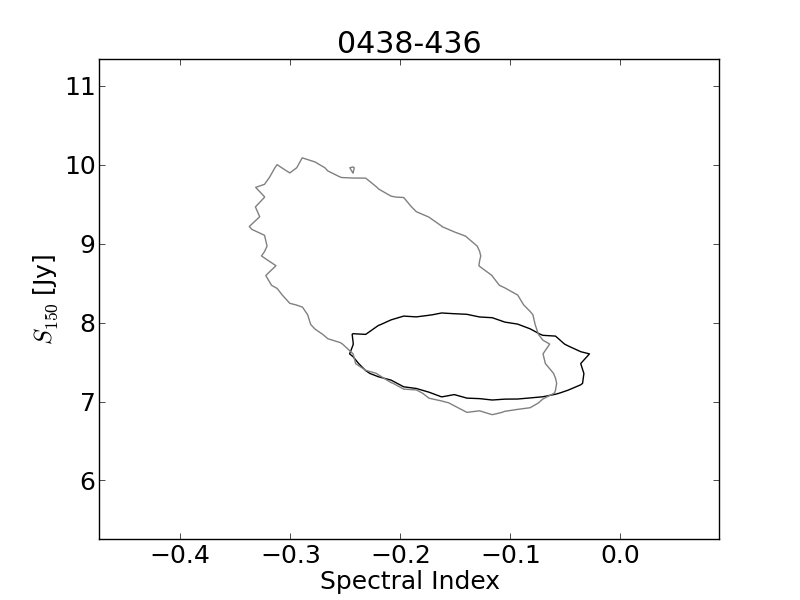} 
\includegraphics[width=2in]{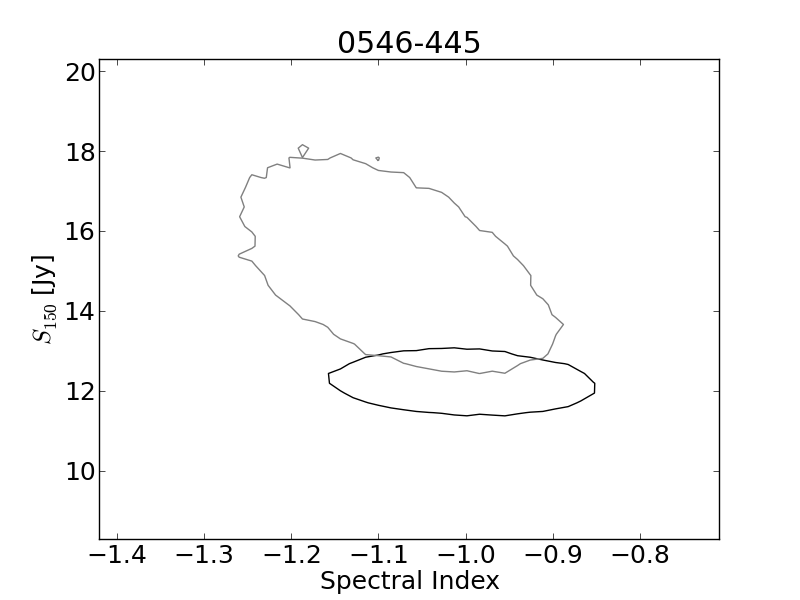} 
\includegraphics[width=2in]{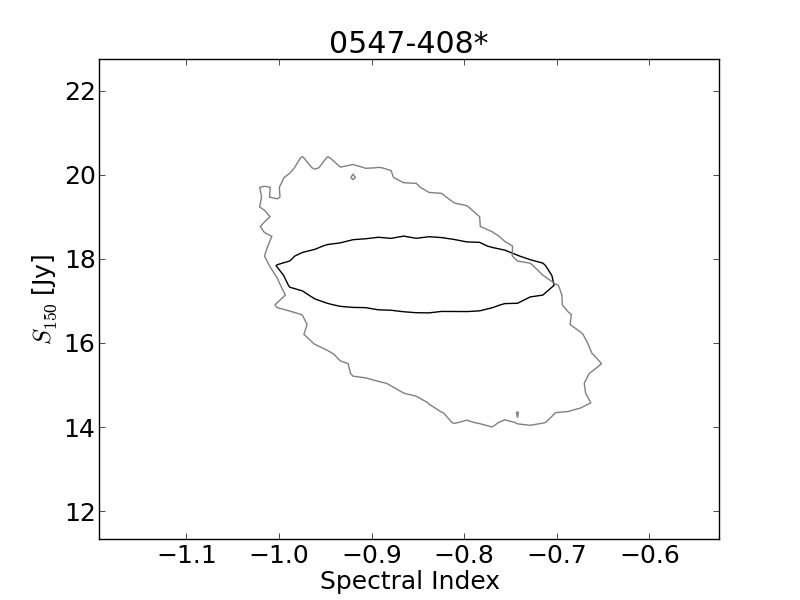} 
\end{center}
\caption{Spectral model contours as described in Figure \ref{fig:pic_spectrum}. 
Catalog fit in grey, black folds in the PAPER 
data. Sources marked with a `*' were used to assess calibration error.
}\label{fig:SI_contour_1}
\end{figure*}

\begin{figure*}[htbp]
\begin{center}
\includegraphics[width=2in]{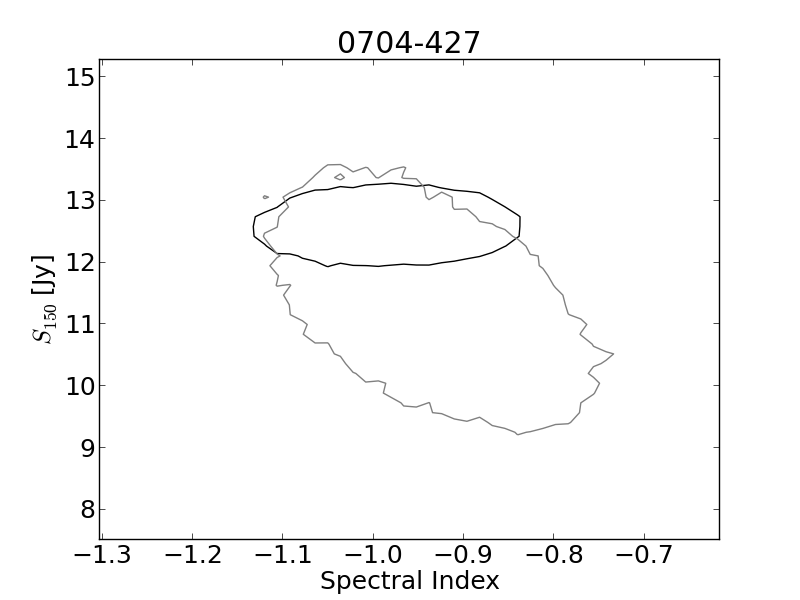} 
\includegraphics[width=2in]{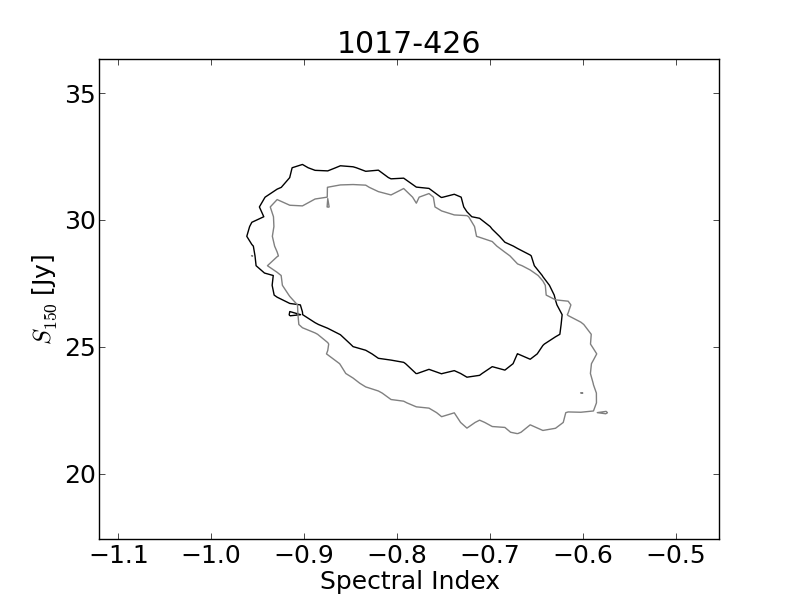} 
\includegraphics[width=2in]{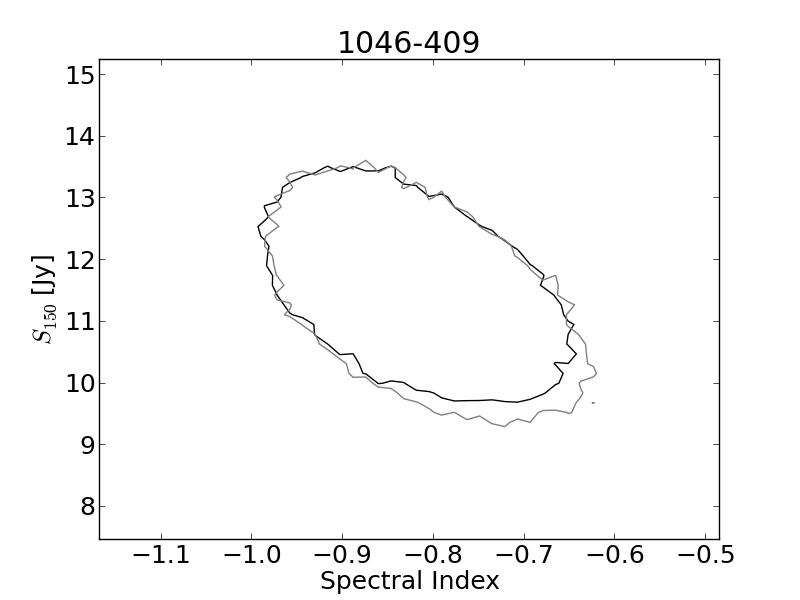} 
\includegraphics[width=2in]{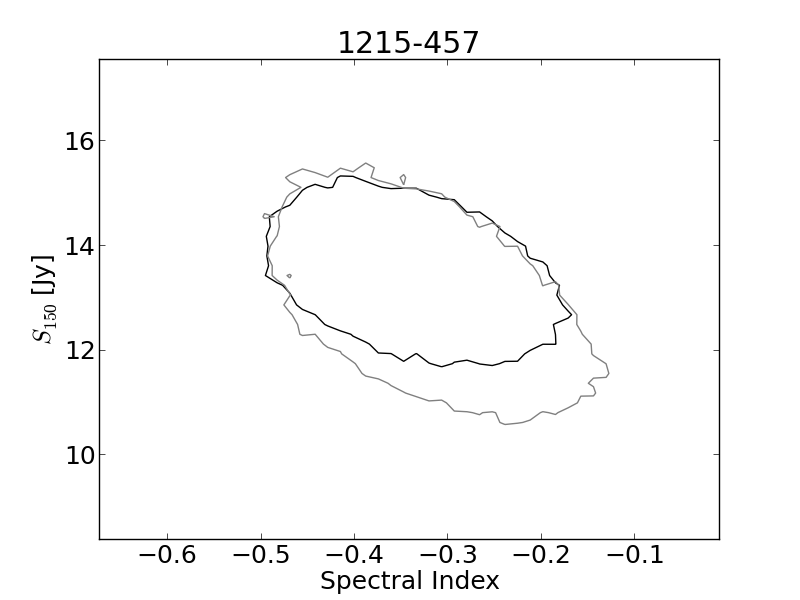} 
\includegraphics[width=2in]{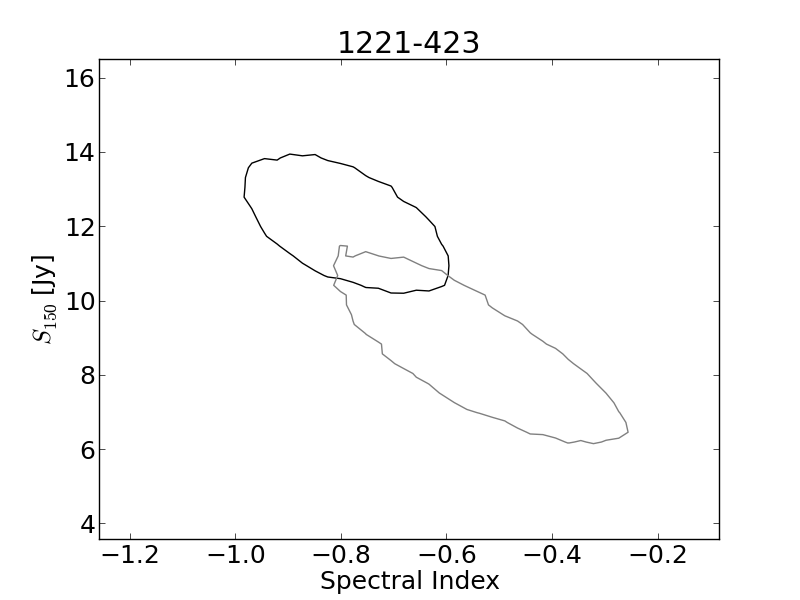} 
\includegraphics[width=2in]{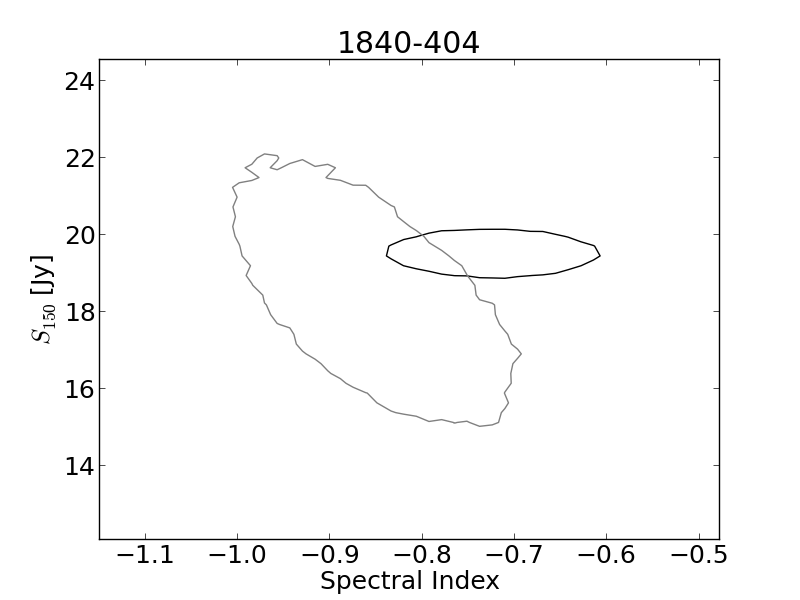} 
\includegraphics[width=2in]{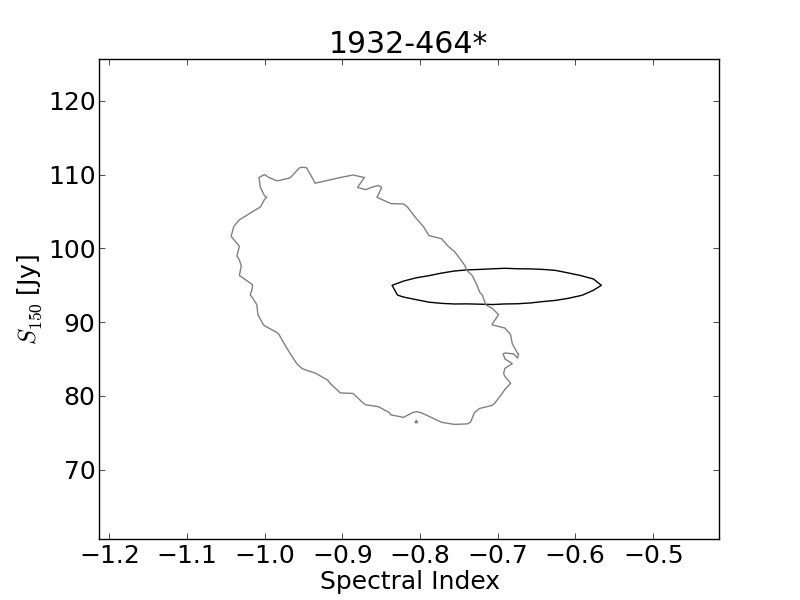} 
\includegraphics[width=2in]{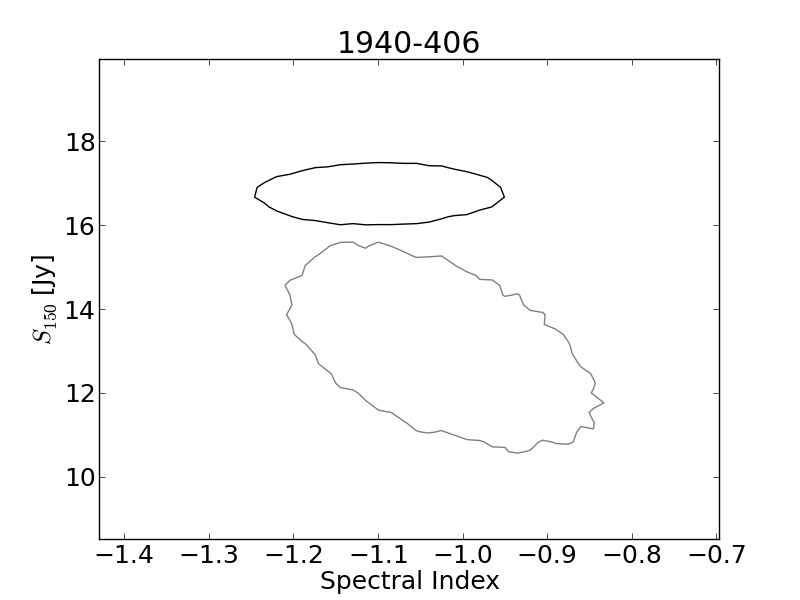} 
\includegraphics[width=2in]{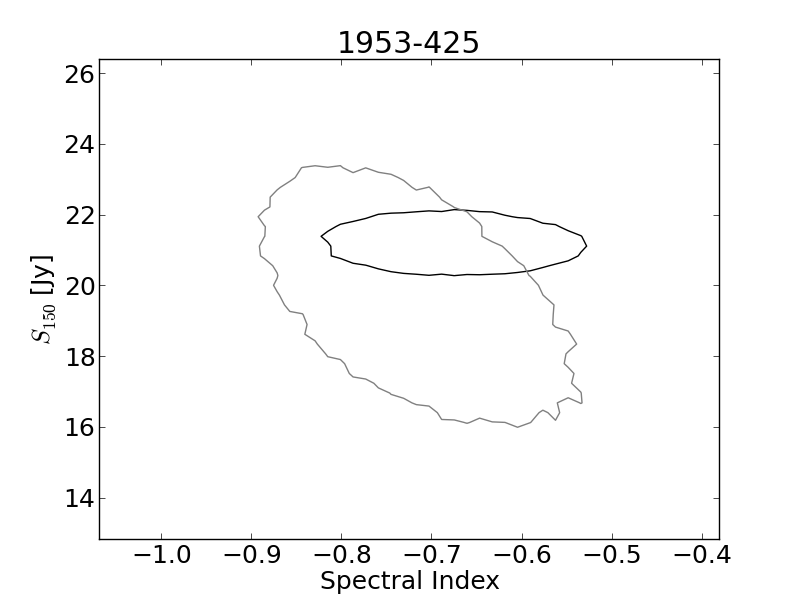} 
\includegraphics[width=2in]{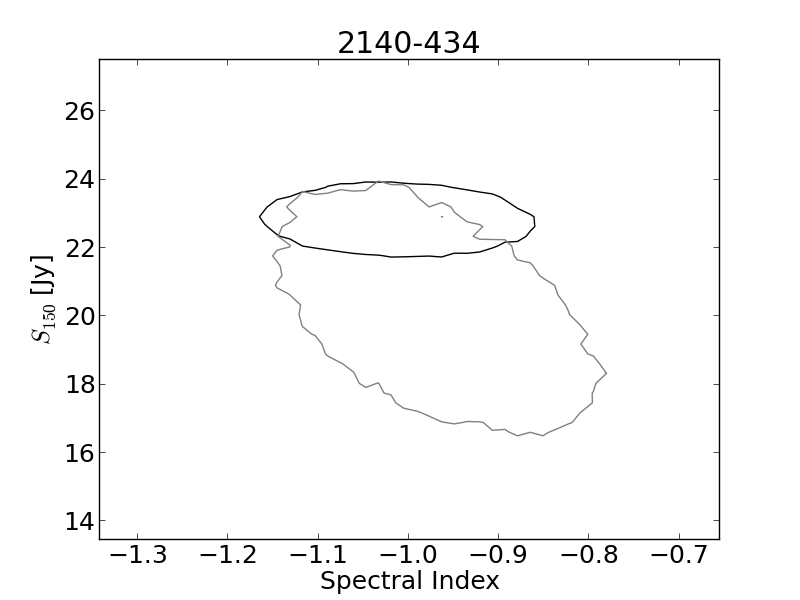} 
\includegraphics[width=2in]{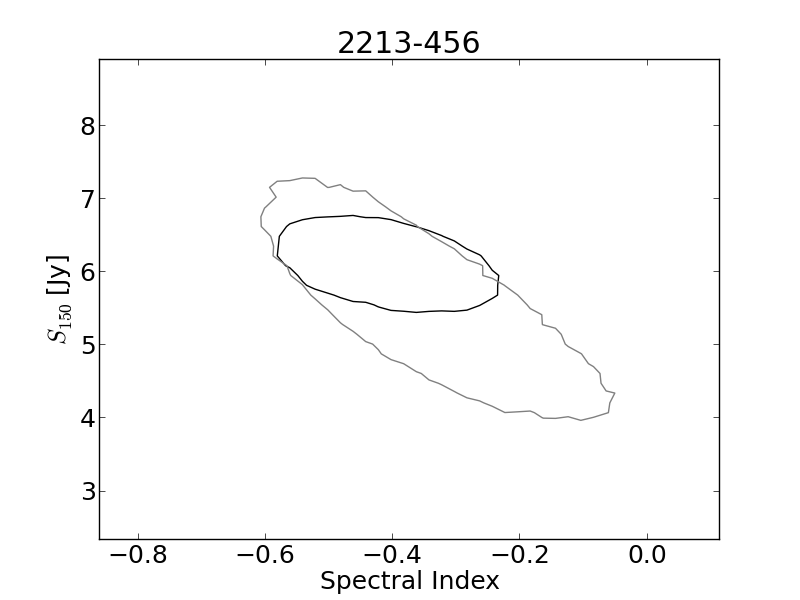} 
\includegraphics[width=2in]{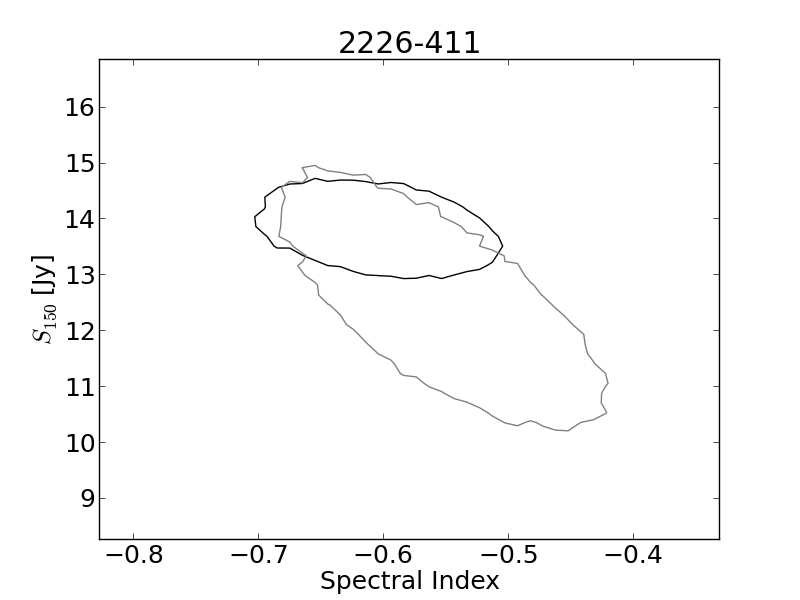} 

\end{center}
\caption{Spectral model contours as described in Figure \ref{fig:pic_spectrum}. 
Catalog fit in grey, black folds in the PAPER 
data.  Sources marked with a `*' were used to assess calibration error.
}\label{fig:SI_contour_2}
\end{figure*}

\begin{figure*}[htbp]
\begin{center}
\includegraphics[width=2in]{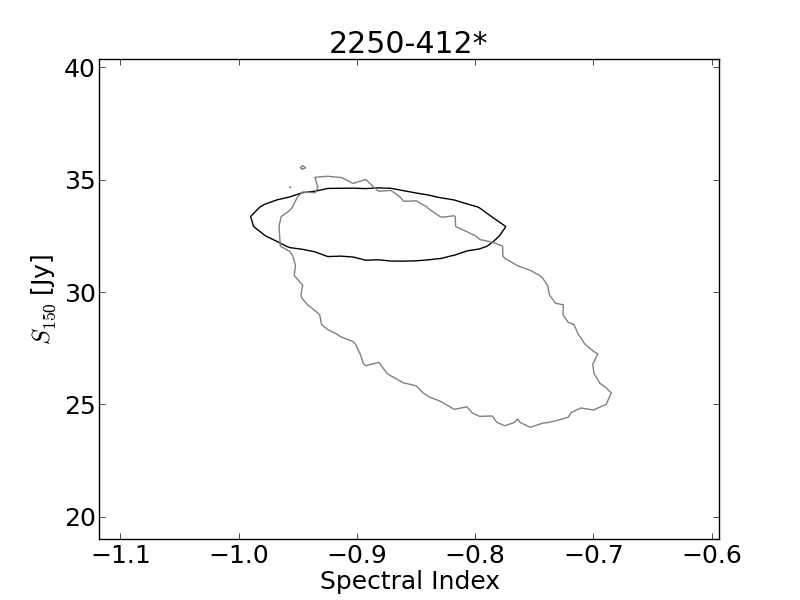} 
\includegraphics[width=2in]{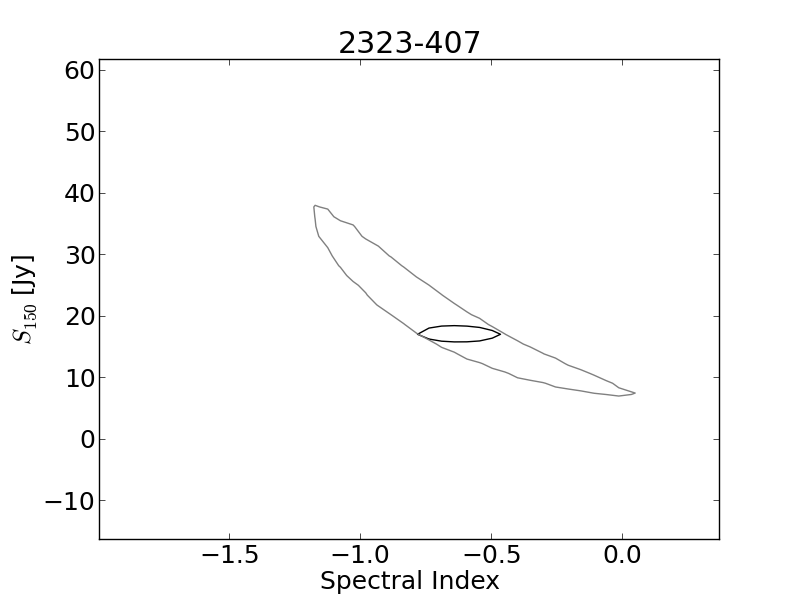} 
\includegraphics[width=2in]{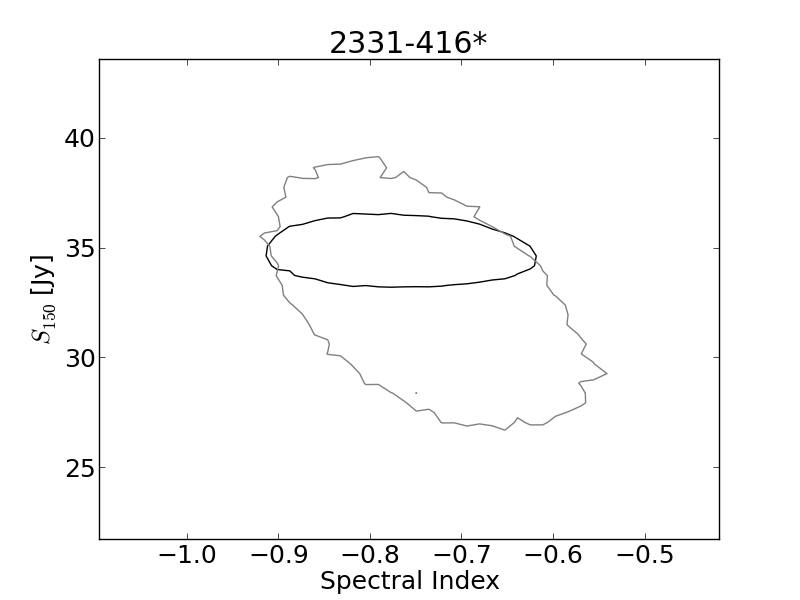} 
\end{center}
\caption{fits of the last three sources, as described in Figure \ref{fig:SI_contour_1}. 
Catalog fit in grey, black folds in the PAPER 
data. 
Sources marked with a
`*' were used to assess calibration error.
}\label{fig:SI_contour_3}
\end{figure*}

However, the vast majority of sources (90\%) have positive improvement index,
indicating strong confirmation of the extrapolated spectrum (see Figures
\ref{fig:SI_contour_1} - \ref{fig:SI_contour_3}).  Pictor A is close to the top of
this group  with an improvement of 0.942 ---only two sources show stronger
confirmation.  The flux model is $S150=$382 $\pm$ 5.4Jy, $\alpha=-0.76\pm0.01$
 a fractional error of 1.4\%.

This represents an order of magnitude improvement in the accuracy of this primary 
flux calibrator. In fact the uncertainty is so small that it pays to consider its reasonableness.
First consider that we have fit the model to many data points, including  7 measurements
(5 PAPER, 2 Wills) accurate to 3\%. If the errors were completely Guassian
the net error would be $3\%/\sqrt(7) = 1.14$\%. Consider also that, where most surveys
might include on the order of ten independent measures of a bright source from different facets to arrive at 
a 5 to 20\% uncertainty,  we have included thousands of independent measurements
over the full horizon to horizon transit, while carefully controlling for systematic variation and resolution affects.

 \section{Conclusion}
 \label{sec:Conclusion}

This paper presents a set of total flux measurements at multiple frequencies within
a 100--200-MHz band, providing
an absolute flux calibration for southern hemisphere EoR studies, as well as a
modest set of verification fluxes for other sources suitable for constraining future primary beam studies.
We have provided a measurement of Pictor A with enough precision to confirm
a linear spectral index between 150 and 600MHz. We apply the same
filtered beamforming method to measure the spectra of bright sources with similar
primary beam response.

The measurements provided
here are the first calibrated, broad-band spectra to cover the EoR band. Existing EoR band measurements
are accurate to 20\% implying a 40\% uncertainty in the absolute power spectrum
level.  The PAPER Pictor A spectrum is found to be accurate to 
$\sim$3\%, a factor of 7 improvement over previous EoR-band measurements.

This uncertainty includes the 
variation in each PAPER measurement ($\sim$1\%), variation between sources and the errors resulting
from extrapolating the \citet{Baars:1977p9678} scale beyond its original range. These last
two are found simultaneously by fitting the spectral extrapolations of several flux calibrators
at once and account for about half of the error. Though past measurements suggested the possibility
of spectral curvature below 200MHz we have found no evidence for this.  With these measurements we are able 
to confirm a single spectral index model for Pictor between 120MHz and 600MHz.

A set of  31 additional verification sources were also targeted to provide additional 
characterization of the flux scale as well as an overall assessment of the catalog accuracy.
Using a Bayesian analysis, we conclude that most of
these are consistent with previous measurements, provide useful new
constraints and support the conclusion that the Pictor A flux is on the correct scale.

Direct measurements of the Pictor A spectrum are key to correctly setting the
flux scale of PAPER, MWA and future EoR experiments like the Hydrogen Epoch of Reionization Array (HERA).
These spectra provide tighter constraints on many of the EoR band fluxes, while
limiting the pernicious effect of primary beam uncertainty.  Future work will use these
fluxes to further refine the primary beam models of these experiments which is crucial to properly reconstructing
both image and power spectrum flux.
Although we have focused on a narrow declination range in this paper, the techniques we describe
may be applied over any declination range.  Future work will also aim to extend this analysis
to other declination ranges, and to tie the flux calibration of these other declination ranges
back to the absolute flux scale we derive here.

\section*{Acknowledgements}

The PAPER project is supported by the National Science Foundation (awards
0804508, 1129258, and 1125558), and a generous grant from the Mt. Cuba
Astronomical Association.
This work makes use of the ``MCMC Hammer" emcee python library \citep[
\url{http://danfm.ca/emcee/}]{ForemanMackey:2012p8684}  and the NASA/IPAC
Extragalactic Database (NED) which is operated by the Jet Propulsion
Laboratory, California Institute of Technology, under contract with the
National Aeronautics and Space Administration.

\clearpage


\begin{deluxetable*}{lllllllllllll}
\tablecolumns{13}
\tablecaption{PAPER spectra for 32 MRC sources}
\tablehead{
\colhead{Name} &
\colhead{\parbox[c][3em]{2em}{Ra \unit{deg}}} & 
\colhead{\parbox[c][2em]{2em}{Dec \unit{deg}}} &
\colhead{\parbox[c][2em]{2em}{S125 \unit{Jy}}} &
\colhead{\parbox[c][2em]{2em}{rms \unit{Jy}}} &
\colhead{\parbox[c][2em]{2em}{S135 \unit{Jy}}} &
\colhead{\parbox[c][2em]{2em}{rms \unit{Jy}}} &
\colhead{\parbox[c][2em]{2em}{S145 \unit{Jy}}} &
\colhead{\parbox[c][2em]{2em}{rms \unit{Jy}}} &
\colhead{\parbox[c][2em]{2em}{S155 \unit{Jy}}} &
\colhead{\parbox[c][2em]{2em}{rms \unit{Jy}}} &
\colhead{\parbox[c][2em]{2em}{S165 \unit{Jy}}} &
\colhead{\parbox[c][2em]{2em}{rms \unit{Jy}}}
}
\startdata
Pictor A&	80.09&	-45.78&	455.3&	13.3&	409.6&	15.8&	389.2&	12.4&	363.5&	11.1&	350.6&	9.1 \\
0003-428&	1.68&	-42.5&	14.5&	1.7&	13.1&	1.4&	12.8&	1.2&	12.6&	1.1&	10.3&	1.0 \\
0007-446&	2.8&	-44.3&	23.2&	2.2&	21.1&	2.2&	20.2&	1.7&	18.6&	1.4&	17.2&	1.6 \\
0008-421&	2.89&	-41.81&	-0.2&	0.8&	-0.3&	1.0&	0.9&	0.9&	0.7&	0.9&	1.3&	1.0 \\
0039-445&	10.7&	-44.16&	29.2&	3.0&	27.6&	2.2&	26.4&	2.2&	24.0&	1.6&	22.2&	1.5 \\
0043-424&	11.73&	-42.05&	48.6&	4.2&	45.2&	3.5&	43.9&	2.3&	40.8&	2.2&	38.4&	2.4 \\
0048-447&	12.87&	-44.4&	16.2&	1.7&	14.6&	1.3&	13.7&	1.6&	13.3&	1.3&	11.3&	1.2 \\
0049-433&	13.22&	-43.03&	20.9&	2.5&	19.0&	2.2&	20.8&	2.2&	19.0&	1.5&	17.4&	1.2 \\
0103-453&	16.48&	-45.02&	31.3&	3.0&	29.0&	2.7&	27.1&	2.9&	25.4&	1.8&	24.3&	2.0 \\
0201-440&	31.05&	-43.76&	13.6&	2.2&	12.6&	1.5&	11.7&	1.0&	11.0&	1.0&	12.0&	1.2 \\
0438-436&	70.17&	-43.53&	9.5&	1.4&	6.0&	0.7&	6.6&	0.6&	5.8&	0.7&	10.5&	1.0 \\

\enddata
\label{tab:data}
\end{deluxetable*}

\begin{deluxetable*}{cccccccccccc}
\tablecolumns{12}
\tablecaption{Spectral fits for 30\tablenotemark{1} MRC sources. Before and after the addition of PAPER data. 90\% agree well with prior
measurements and demonstrate increased precision over previous measurements. Full table available online.}
\tablehead{
\multicolumn{3}{c}{ }&
\multicolumn{4}{c}{PAPER + Catalog}&
\multicolumn{4}{c}{Catalog\tablenotemark{2}}&\\
\colhead{Name} &
\colhead{\parbox[c][3em]{2em}{Ra\\ \unit{deg}}}& 
\colhead{\parbox[c][3em]{2em}{Dec\\ \unit{deg}}} &
\colhead{\parbox[c][3em]{2em}{$S150$\\ \unit{Jy}}} &
\colhead{\parbox[c][3em]{2em}{$\Delta$S\\ \unit{Jy}}} &
\colhead{\parbox[c][3em]{2em}{$\alpha$\\ \unit{--}}} &
\colhead{\parbox[c][3em]{2em}{$\Delta\alpha$\\ \unit{--}}} &
\colhead{\parbox[c][3em]{2em}{$S150_p$\\ \unit{Jy}}} &
\colhead{\parbox[c][3em]{2em}{$\Delta S_p$\\ \unit{Jy}}} &
\colhead{\parbox[c][3em]{2em}{$\alpha_p$\\ \unit{--}}} &
\colhead{\parbox[c][3em]{2em}{$\Delta\alpha_p$\\ \unit{--}}} &
\colhead{\parbox[c][3em]{2em}{Imp.\tablenotemark{3}\\ \unit{--}}}
}
\startdata
Pictor A&	80.09&	-45.78&	381.88&	5.36&	-0.76&	0.01&	392.63&	21.18&	-0.77&	0.04&	0.942 \\
0003-428&	1.68&	-42.5&	12.24&	0.61&	-0.86&	0.11&	12.42&	2.75&	-0.86&	0.19&	0.636 \\
0007-446&	2.8&	-44.3&	19.17&	0.82&	-1.02&	0.11&	18.08&	4.0&	-0.98&	0.19&	0.704 \\
0039-445&	10.7&	-44.16&	24.78&	0.91&	-0.86&	0.12&	22.6&	4.96&	-0.78&	0.19&	0.801 \\
0043-424&	11.72&	-42.05&	41.69&	1.27&	-0.77&	0.1&	36.14&	4.93&	-0.69&	0.12&	0.34 \\
0048-447&	12.87&	-44.4&	13.24&	0.64&	-0.99&	0.12&	13.37&	2.81&	-0.99&	0.19&	0.657 \\
0049-433&	13.22&	-43.03&	18.99&	0.84&	-0.91&	0.11&	21.2&	4.58&	-1.01&	0.2&	0.754 \\
0103-453&	16.48&	-45.02&	25.73&	1.07&	-1.11&	0.1&	21.75&	2.85&	-1.04&	0.12&	0.177 \\
0201-440&	31.05&	-43.76&	11.71&	0.66&	-0.6&	0.12&	10.19&	6.72&	-0.5&	0.44&	2.805 \\
0438-436&	70.17&	-43.53&	7.51&	0.38&	-0.14&	0.08&	8.35&	1.13&	-0.21&	0.1&	0.305 \\

\enddata
\label{tab:fits}
\tablenotetext{1}{0008-421 is self-absorbed at 150MHz and is not listed here, while 1 did not have sufficient catalog data
for a spectral fit and was also excluded}
\tablenotetext{2}{MCMC fits to prior catalog data, before addition of PAPER measurements}
\tablenotetext{3}{SED figure of merit change times confidence overlap, a measure of accuracy and precision.  Higher
value indicates an increase in both model fit precision and accuracy.}
\end{deluxetable*}

\end{document}